\documentclass[]{aa}  
\usepackage{booktabs}
\usepackage{graphicx}
\usepackage{verbatim}
\usepackage{breqn}
\usepackage{txfonts}
\usepackage{subfigure}
\usepackage{gensymb}
\usepackage{natbib,twoopt}
\usepackage[breaklinks=true,colorlinks=true, urlcolor=blue,linkcolor=red]{hyperref} %% to avoid \citeads line fills

\newcommand{\nustar}{\textit{NuSTAR}}
\newcommand{\chandra}{\textit{Chandra}}
\newcommand{\suzaku}{\textit{Suzaku}}
\newcommand{\xmm}{\textit{XMM-Newton}}
\graphicspath{{/media/alberto/TRANSCEND/mrk1210/figures/},{/Volumes/TRANSCEND/mrk1210/figures/}}
\bibpunct{(}{)}{;}{a}{}{,} %% natbib format for A&A and ApJ
\makeatletter
\newcommandtwoopt{\citeads}[3][][]{\href{http://adsabs.harvard.edu/abs/#3}%
{\def\hyper@linkstart##1##2{}%
\let\hyper@linkend\@empty\citealp[#1][#2]{#3}}}

\newcommandtwoopt{\citepads}[3][][]{\href{http://adsabs.harvard.edu/abs/#3}%
{\def\hyper@linkstart##1##2{}%
\let\hyper@linkend\@empty\citep[#1][#2]{#3}}}

\newcommandtwoopt{\citetads}[3][][]{\href{http://adsabs.harvard.edu/abs/#3}%
{\def\hyper@linkstart##1##2{}%
\let\hyper@linkend\@empty\citet[#1][#2]{#3}}}

\newcommandtwoopt{\citeyearads}[3][][]%
{\href{http://adsabs.harvard.edu/abs/#3}
{\def\hyper@linkstart##1##2{}%
\let\hyper@linkend\@empty\citeyear[#1][#2]{#3}}}
\makeatother

\begin{document} 

\title{The Phoenix galaxy as seen by \nustar}

\author{A. Masini\inst{1,2}, A. Comastri\inst{1}, S. Puccetti\inst{3,4}, M. Balokovi\'c\inst{5}, P. Gandhi\inst{6,7}, M. Guainazzi\inst{8,9}, F.\,E. Bauer\inst{10,11,12,13}, \\ S.\,E. Boggs\inst{14}, P.\,G. Boorman\inst{7}, M. Brightman\inst{5},  F.\,E. Christensen\inst{15}, W.\,W. Craig\inst{14,16}, D. Farrah\inst{17}, C.\,J. Hailey\inst{18}, \\ F.\,A. Harrison\inst{5}, M.\,J. Koss\inst{19}, S.\,M. LaMassa\inst{20}, C. Ricci\inst{10,13}, D. Stern\inst{21}, D.\,J. Walton\inst{21,22} and W.\,W. Zhang\inst{20}}
 %1
\institute{INAF-Osservatorio Astronomico di Bologna, via Ranzani 1, 40127 Bologna, Italy\\ \email{\href{mailto:alberto.masini4@unibo.it}{alberto.masini4@unibo.it}}
         \and 
%2
Dipartimento di Fisica e Astronomia (DIFA), Universit\`a  di Bologna, viale Berti Pichat 6/2, 40127 Bologna, Italy
         \and 
%3 
ASDC-ASI, Via del Politecnico, 00133 Roma, Italy
        \and 
%4
INAF-Osservatorio Astronomico di Roma, Via Frascati 33, 00040 Monte Porzio Catone, Italy
	\and 
%5
Cahill Center for Astronomy and Astrophysics, California Institute of Technology, Pasadena, CA 91125, USA
	\and 
%6
Centre for Extragalactic Astronomy, Department of Physics, University of Durham, South Road, Durham DH1 3LE, UK
        \and 
%7
School of Physics and Astronomy, University of Southampton, Highfield, Southampton SO17 1BJ, UK
	\and 
%8
Institute of Space and Astronatical Science (JAXA), 3-1-1 Yoshinodai, Sagamihara, Kanagawa, 252-5252, Japan
	\and
%9
European Space Astronomy Center of ESA, P.O.Box 78, Villanueva de la Ca\~nada, E-28691 Madrid, Spain
	\and
%10	
Instituto de Astrof{\'{\i}}sica and Centro de Astroingenier{\'{\i}}a, Facultad de F{\'{i}}sica, Pontificia Universidad Cat{\'{o}}lica de Chile, Casilla 306, Santiago 22, Chile
	\and
%11
Millennium Institute of Astrophysics (MAS), Nuncio Monse{\~{n}}or S{\'{o}}tero Sanz 100, Providencia, Santiago, Chile
	\and
%12
Space Science Institute, 4750 Walnut Street, Suite 205, Boulder, Colorado 80301
	\and
%13
EMBIGGEN Anillo, Concepci\'on, Chile
	\and
%14
Space Science Laboratory, University of California, Berkeley, CA 94720, USA
	\and
%15
DTU Space National Space Institute, Technical University of Denmark, Elektrovej 327, 2800 Lyngby, Denmark
	\and
%16
Lawrence Livermore National Laboratory, Livermore, CA 94550, USA
	\and
%17
Department of Physics,  Virginia Tech,  Blacksburg,  VA 24061, USA
	\and 
%18
Columbia Astrophysics Laboratory, Columbia University, New York, NY 10027, USA
	\and
%19
Institute   for   Astronomy,   Department   of   Physics,   ETH Zurich,  Wolfgang-Pauli-Strasse  27,  CH-8093  Zurich,  Switzerland
	\and
%20
NASA Goddard Space Flight Center, Greenbelt, MD 20771, USA
	\and 
%21
Jet Propulsion Laboratory, California Institute of Technology, Pasadena, CA 91109, USA
	\and 
%22
Space Radiation Laboratory, California Institute of Technology, Pasadena, CA 91125, USA}

%\date{Received September 15, 1996; accepted March 16, 1997}

% \abstract{}{}{}{}{} 
% 5 {} token are mandatory
\abstract
   {}
   {We study the long-term variability of the well-known  Seyfert\,2 galaxy Mrk\,1210 (a.k.a. UGC\,4203, or the Phoenix galaxy).}
   {The source was observed by many X-ray facilities in the last 20 years. Here we present a \nustar\ observation and put the results in context of previously published observations.}
   {\nustar \ observed Mrk\,1210 in 2012 for 15.4 ks. The source showed Compton-thin obscuration similar to that observed by \chandra, \suzaku, \textit{BeppoSAX} and \xmm \ over the past two decades, but different from the first observation by \textit{ASCA} in 1995, in which the active nucleus was caught in a low flux state -- or obscured by Compton-thick matter, with a reflection-dominated spectrum. Thanks to the high-quality hard X-ray spectrum obtained with \nustar\ and exploiting the long-term spectral coverage spanning 16.9 years, we can precisely disentangle the transmission and reflection components and put constraints on both the intrinsic long-term variability and hidden nucleus scenarios. In the former case, the distance between the reflector and the source must be at least $\sim$ 2 pc, while in the latter one the eclipsing cloud may be identified with a water maser-emitting clump.}
   {}


   \titlerunning{\nustar \ observation of Mrk\,1210}
   \authorrunning{A. Masini et al.} 
   \maketitle
%
%________________________________________________________________

\section{Introduction}
\begin{figure}
\centering
\includegraphics[width=0.3\textwidth]{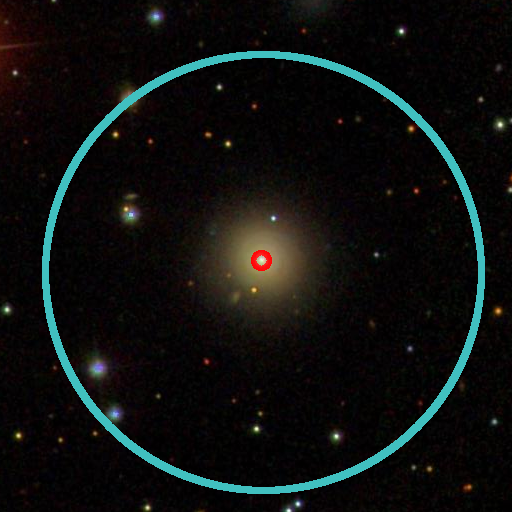}
\caption{Sloan Digital Sky Survey (SDSS) optical image of Mrk\,1210 (gri composite). The cyan and red circles are the source extraction regions for the \nustar\ (see \S \ref{sec:data}) and \chandra\ (see Appendix) spectra, respectively.}
\label{fig:optical}
\end{figure}

X-ray variability is a well-known property of active galactic nuclei (AGN). In recent years, many studies have focused on its characterization, and a class of extremely variable sources was found. Sources showing a transition between Compton-thin (i.e., with an obscuring column density of $10^{22} < N_{\rm H} < 10^{24}$ cm$^{-2}$) and Compton-thick levels of obscuration ($N_{\rm H} > 10^{24}$ cm$^{-2}$) are called ``changing-look AGN'', and are important to assess the relevance and physics of variability. The most famous cases are NGC\,1365 \citepads{2005ApJ...623L..93R,2014ApJ...788...76W,2015ApJ...804..107R}, NGC\,6300 \citepads{2002MNRAS.329L..13G}, NGC\,2992 \citepads{2000A&A...355..485G} , NGC\,7674 (\citeads{2005A&A...442..185B}, Gandhi et al. submitted), and NGC\,7582 \citepads{2007A&A...466..855P,2015ApJ...815...55R}. The nature of X-ray variability could be explained in different ways. A drop in flux can either be due to intrinsic fading of the central engine since we do not expect the accretion of matter on supermassive black holes (SMBHs) to be constant in time, or to an eclipsing phenomenon caused by some clumpy material absorbing the radiation along the line of sight (l.o.s.). There could be also other effects, like instabilities into the accretion flow. A significant change in the spectral shape with a constant flux can also occur. \newline 
In order to adequately study this complex property of AGN, monitoring sources on a wide range of timescales, from weeks to years is needed, ideally with high sensitivity in the hard X-ray band (> 10 keV), which more directly probes the primary emission from the innermost regions of the AGN. \newline 
Mrk\,1210 ($z = 0.0135$, Figure \ref{fig:optical}), also known as UGC\,4203, hosts a Seyfert\,2 AGN which was initially observed by \textit{ASCA} in 1995 \citepads{2000ApJ...542..175A}. The flat spectrum and high equivalent width (EW) of the iron line at 6.4 keV were interpreted as emerging from reflection off circumnuclear matter of the AGN primary X-ray continuum, severely suppressed by Compton-thick absorption. In 2001, an \xmm\ observation \citepads{2002A&A...388..787G} found that Mrk\,1210, six years after the first observation, was still obscured but at the Compton-thin level only. Interpreting the change as an intrinsic flux enhancement, they coined the name ``Phoenix galaxy'' for this new changing-look AGN. Also \citetads{2004PASJ...56..425O}, in the same year, observed Mrk\,1210 with \textit{BeppoSAX} and found similar results. Later on, the Phoenix galaxy was observed by other instruments, always showing variability of less than a factor of two in both intrinsic emission and column density, with the latter always in the Compton-thin regime \citepads{2009A&A...496..653M,2010MNRAS.406L..20R}. The list of all X-ray observations is presented in Table \ref{tab:obs}. \newline
While the source has been extensively studied for the past twenty years, it is still not clear what is the reason for the changes observed between 1995 and 2001. A definitive answer is not yet available; however, as time went by, a change in $N_{\rm H}$ obscuring the nucleus was progressively addressed as the principal effect causing the change in flux. A better understanding of this source can be achieved thanks to the \textit{Nuclear Spectroscopic Telescope Array} (\nustar). Launched in 2012, \nustar\ is the first telescope able to focus hard X-ray photons, enabling a gain of a factor $\sim 100$ in sensitivity and one order of magnitude in angular resolution with respect to previous facilities in the hard (> 10 keV) X-ray band \citepads{2013ApJ...770..103H}. In its operating band (3--79 keV), thanks to a field of view (FoV) at 10 keV of 10\arcmin~ $\times$ 10\arcmin, and an 18\arcsec~ FWHM with a half-power diameter of 58\arcsec, \nustar \ is suitable for studying the hard X-ray spectra of AGN with high sensitivity. In this paper, we report on the \nustar \ observation of Mrk\,1210 in 2012. After describing the data reduction (\S \ref{sec:data}), we present the spectral analysis in \S \ref{sec:spec}. A discussion of our results in the context of the previous findings in the literature is provided in \S \ref{sec:longterm}, and we draw our conclusions in \S \ref{sec:conclusions}.
\begin{table}
\caption{\label{tab:obs} History of X-ray observations of Mrk\,1210.}
\centering
\begin{tabular}{lc}
\hline\hline
\noalign{\vskip 0.5mm}   
Instrument & Date of observation(s)  \\ \noalign{\vskip 0.5mm} 
\noalign{\vskip 0.5mm}  
\hline
\noalign{\vskip 1mm}  
\textit{ASCA} & 1995-Oct-18, 1995-Nov-12  \\ \noalign{\vskip 0.5mm}  
\xmm \ & 2001-May-5  \\ \noalign{\vskip 0.5mm}  
\textit{BeppoSAX}  & 2001-May-5  \\ \noalign{\vskip 0.5mm} 
\chandra\ & 2004-Mar-4 \\ \noalign{\vskip 0.5mm}
\textit{Swift} & 2006-Oct-6...2008-Apr-29 (4 times) \\ \noalign{\vskip 0.5mm}
\suzaku \ &  2007-May-2 \\ \noalign{\vskip 0.5mm}  
\chandra \ &  2008-Jan-15...2008-Feb-06 (5 times) \\ \noalign{\vskip 0.5mm}  
\nustar \ & 2012-Oct-5 \\ 
\noalign{\vskip 1mm}  
\hline
\end{tabular}
\end{table}
\section{Data reduction}
\label{sec:data}
\nustar \ observed Mrk\,1210 on 2012 October 5 for 15.4 ks. The raw event file was processed using the \nustar \ Data Analysis Software package v. 1.5.1 (NuSTARDAS)\footnote{http://heasarc.gsfc.nasa.gov/docs/nustar/analysis/nustar\_swguide.pdf}. Calibrated and cleaned event files were produced using the calibration files in the \nustar \ CALDB (20150316) and standard filtering criteria with the \texttt{nupipeline} task.  We used the \texttt{nuproducts} task included in the NuSTARDAS package to extract the \nustar \ source and background spectra using the appropriate response and ancillary files. We extracted the source spectrum and light curve in both focal plane modules, FPMA and FPMB, using 87\arcsec-radius circular apertures, while the background was extracted using three source-free circular regions on the same chip as the source. All spectra were binned to a minimum of 20 photons per bin using the HEAsoft task \texttt{grppha}.
\begin{figure}
\centering
\includegraphics[width=0.5\textwidth]{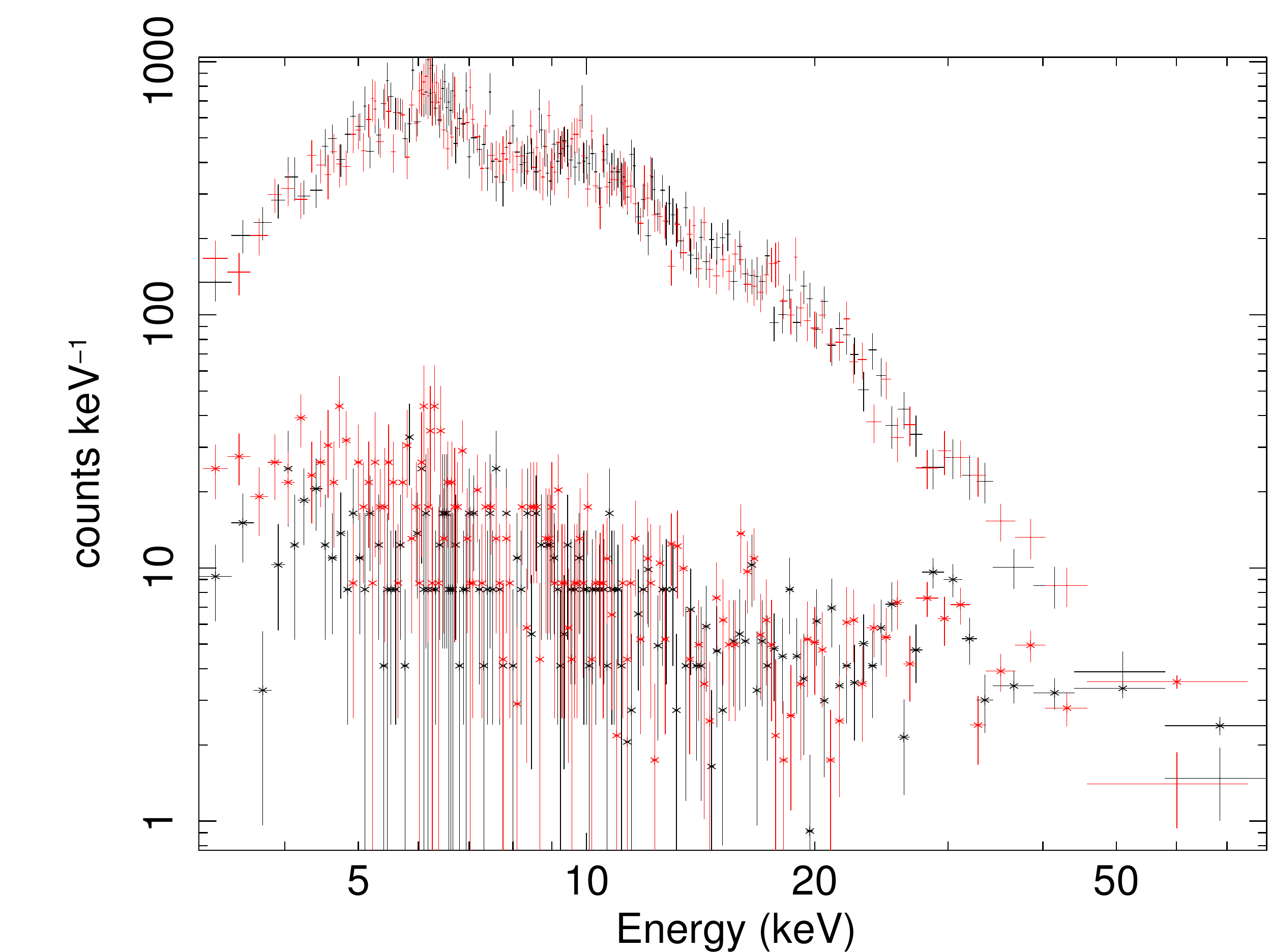}
\caption{\nustar \ background-subtracted spectrum of Mrk\,1210 (FPMA in black, FPMB in red). The background is shown by diamonds, and is lower than the source signal up to $\sim$ 50 keV. Both have been rebinned for plotting clarity.}
\label{fig:spectrum}
\vspace{2mm}
\includegraphics[width=0.45\textwidth]{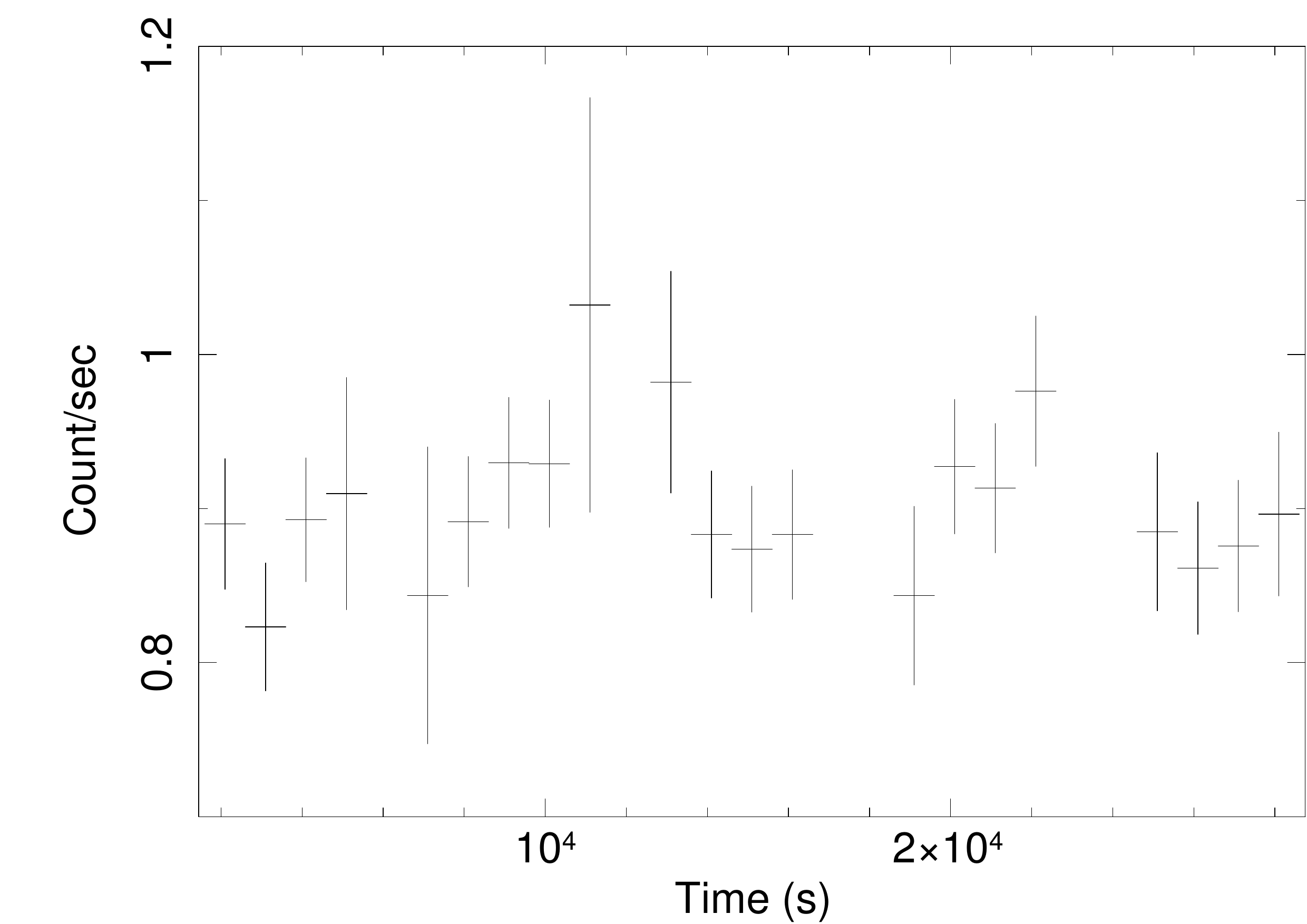}
\caption{3-79 keV lightcurve of Mrk\,1210. The background-subtracted FPMA and FPMB lightcurves were summed.}
\label{fig:lcurve}
\end{figure}
\section{Spectral analysis}
\label{sec:spec}
We used the XSPEC software version 12.9.0 \citepads{1996ASPC..101...17A} to carry out the spectral analysis. The source is clearly detected by \nustar \ up to $\sim$ 60 keV (Figure \ref{fig:spectrum}), with a net count rate of $0.3894 \pm 0.0052$ cts s$^{-1}$ and $0.3762 \pm 0.0052$ cts s$^{-1}$ for FPMA and FPMB, respectively. During the observation, the source kept a nearly constant flux, with amplitude variations of less than 30\% and no systematic trend, as shown in Figure \ref{fig:lcurve}.  \newline
We included in all our fits a Galactic column density $N_{\rm H,gal} = 3.45 \times 10^{20}$ cm$^{-2}$ \citepads{2005A&A...440..775K}. 
\subsection{Phenomenological models}
\begin{figure}
\centering
%\includegraphics[width=0.45\textwidth]{powlaw}
%\caption{Fit with a power law to the \nustar \ spectrum of Mrk\,1210 (FPMA in black, FPMB in red).}
%\label{fig:powlaw}
%\includegraphics[width=0.45\textwidth]{plcabs}
%\caption{Fit with the \texttt{plcabs} + \texttt{zgauss} model to the \nustar \ spectrum of Mrk\,1210 (FPMA in black, FPMB in red).}
%\label{fig:plcabs}
\includegraphics[width=0.44\textwidth]{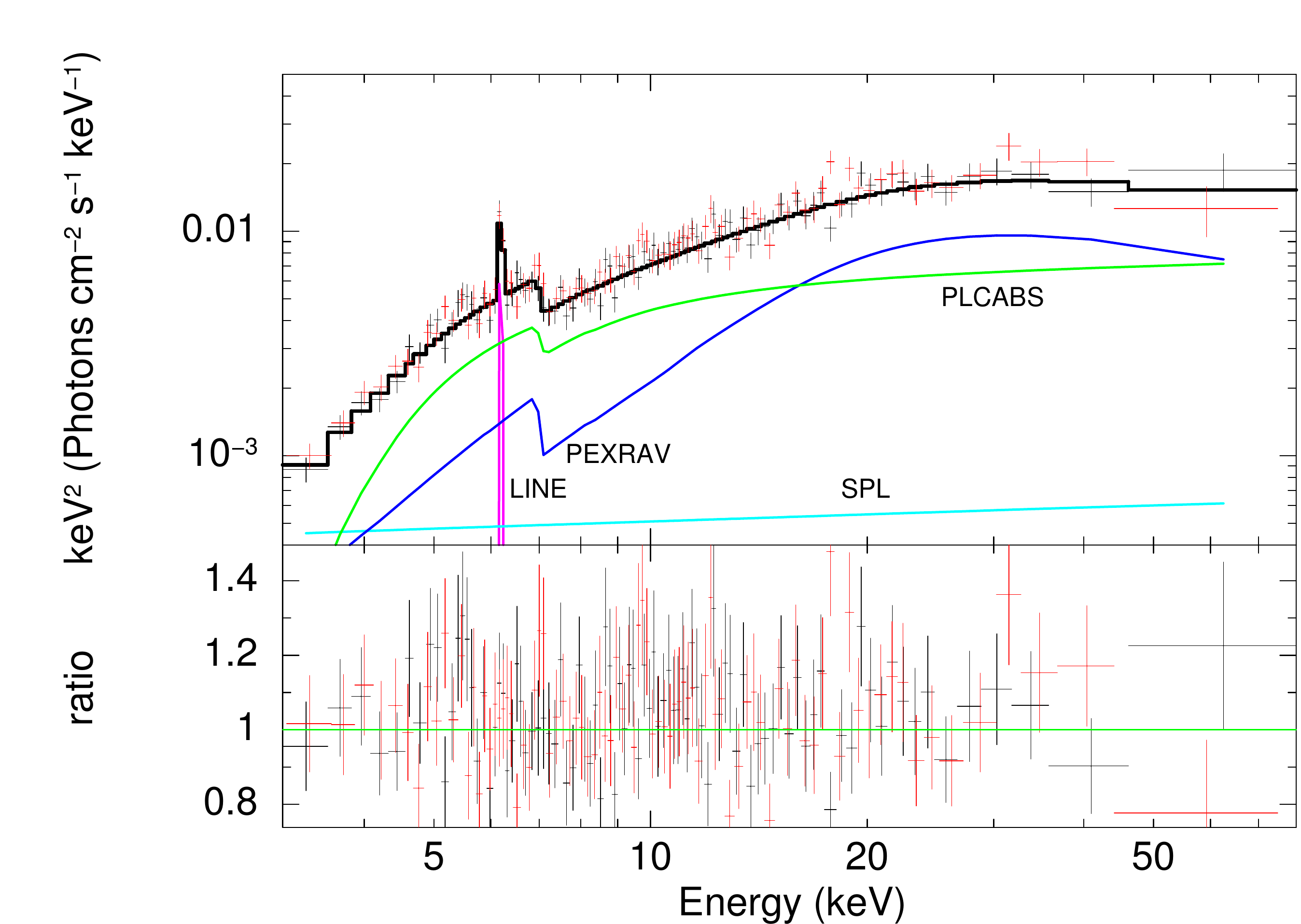}
\caption{Fit with the baseline model \eqref{eq:baseline} to the \nustar \ spectrum of Mrk\,1210 (FPMA in black, FPMB in red).}
\label{fig:baseline}
\vspace{5mm}
\includegraphics[width=0.44\textwidth]{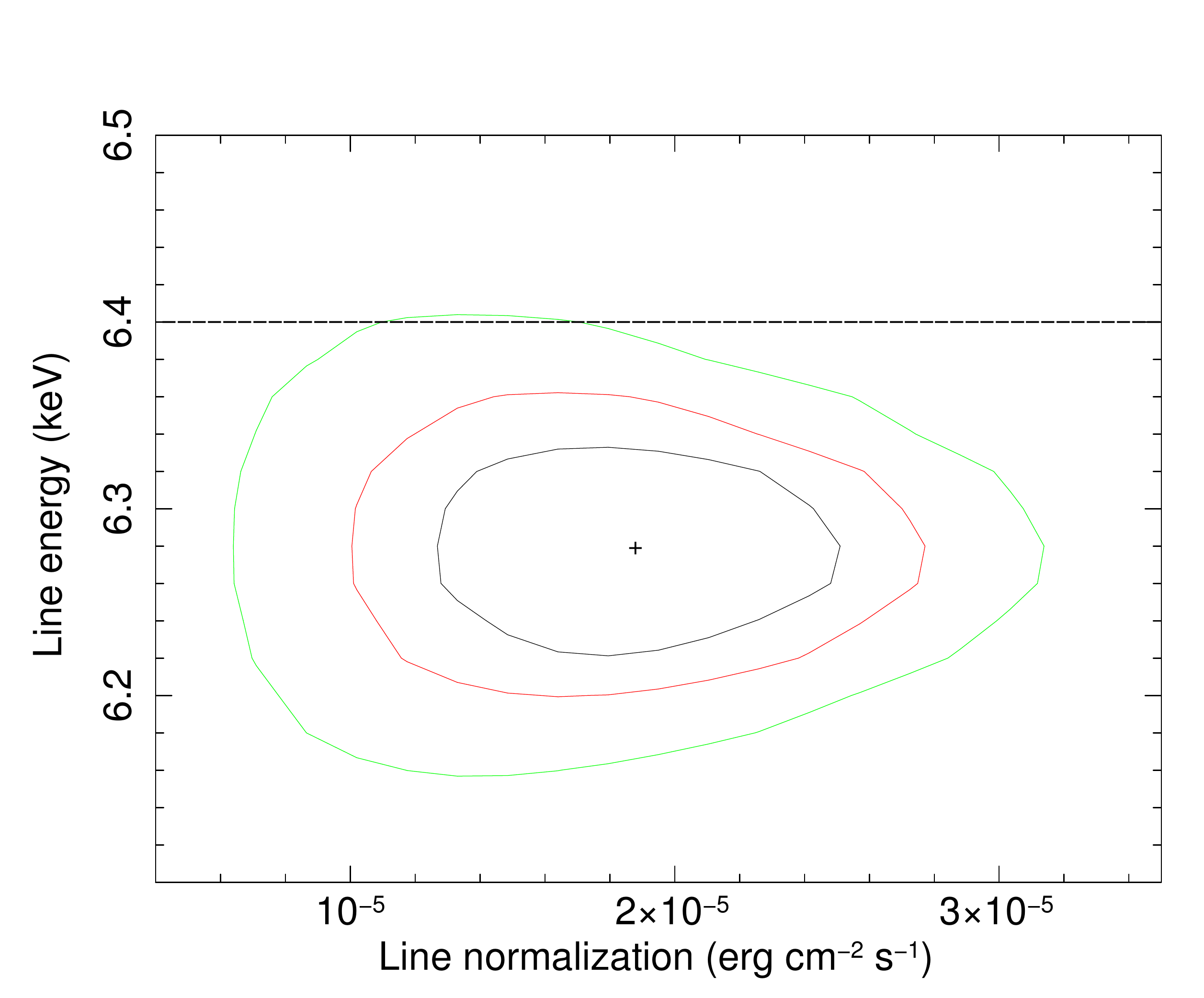}
\caption{68\% (black), 90\% (red), 99\% (green) confidence contours. The two-dimensional confidence contours allow a marginal consistency with the iron K$\alpha$ line at 6.4 keV, indicated with the dashed horizontal line.}
\label{fig:steppar}
%\includegraphics[width=0.45\textwidth]{torus}
%\caption{Fit with the BNTorus model to the \nustar \ spectrum of Mrk\,1210 (FPMA in black, FPMB in red).}
%\label{fig:torus}
\end{figure}

Since we know from previous work that a simple Galactic-absorbed power law is not able to reproduce the spectral complexity of Mrk\,1210, we started fitting its spectrum with a \texttt{plcabs} model, a power law with an exponential cutoff, taking into account Compton scattering and absorption at the source \citepads{1997ApJ...479..184Y}. We also added a narrow ($\sigma = 10$ eV, fixed) Gaussian line component to fit the clearly visible feature at $\sim$ 6 keV.
 The fit was good ($\chi^{2}/\nu = 537/480$), with a photon index $\Gamma = 1.40 \pm 0.05$ and an obscuring column of gas along the l.o.s. of $N_{\rm H} = (1.7 \pm 0.2) \times 10^{23}$ cm$^{-2}$. From previous studies, an additional soft component in the X-ray spectrum of the Phoenix galaxy is known \citepads[e.g.,]{2002A&A...388..787G}. Since this component can contribute between 3 and $\sim$ 5 keV in the \nustar\ band, we modeled it adding another power law, and linking all the parameters (photon index, redshift, normalization) to the ones of the \texttt{plcabs} component. We then allowed for a constant multiplicative value (referred to as $f_{\rm s}$) to vary in the fit, which represents the fraction of the X-ray continuum Thomson-scattered into the soft X-rays. This procedure is often employed in the literature, and fractions of a few percent are typical  \citepads{2007AIPC..924..822B}. Adding a such scattered power law (SPL) improved the fit ($\chi^{2}/\nu = 521/479$) at more than a 99\% confidence limit, based on an F-test. \newline
Although the fit was already acceptable, we wanted to test if any reflection component was required by the data. We therefore added a \texttt{pexrav} model \citepads{1995MNRAS.273..837M}, which includes Compton reflection on a slab of neutral material with infinite optical depth. The final model in XSPEC notation is the following:
\begin{equation}
\label{eq:baseline}
\begin{split}
\mathrm{BASELINE} =\overbrace{ \mathrm{constant}}^{\text{cross-normalization}}\times\overbrace{\mathrm{phabs}}^{\text{Gal. absorption}}\times \\
\times~ (\overbrace{ \mathrm{plcabs} + \mathrm{pexrav} + \mathrm{zgauss} }^{\text{nuclear emission}}~ +\\
  +\underbrace{ f_{\rm s}\times\mathrm{zpowerlw}}_{\text{soft component (SPL)}}).
\end{split}
\end{equation}
We will refer to this as the ``baseline'' model. The fit improved dramatically ($\chi^{2}/\nu = 478/478$, $\Delta\chi^{2} = 43$, see Figure \ref{fig:baseline}): the source had a photon index $\Gamma = 1.9 \pm 0.1$, a column density $N_{\rm H} = 3.0^{+ 0.7}_{- 0.6} \times 10^{23}$ cm$^{-2}$, a reflection parameter $R = 2.5^{+ 1.2}_{- 0.9}$ (intended as the ratio of the \texttt{pexrav} normalization and the \texttt{plcabs} one) and a fraction of the primary power law scattered in the soft part of the spectrum of $8^{+ 5}_{- 6}\%$, which is higher than the average of Seyfert\,2 galaxies, but not unusual \citepads{2007AIPC..924..822B}. However, this latter component is now only significant at the 97\% confidence limit, due to the high level of reflection. Removing it, the fit gets slightly worse ($\chi^{2}/\nu = 483/479$), but the parameters are the same within the uncertainties. \newline We also note that the best fit line energy is lower than the iron K$\alpha$ 6.4 keV centroid, although consistent with it at a 99\% confidence limit (Figure \ref{fig:steppar}). Using  previous versions of the of the NuSTARDAS software (v 1.2.1) and \nustar \ CALDB (20130909), and applying the same baseline model, we find that all the fit parameters are the same within the uncertainties, with a centroid line energy of $6.35 \pm 0.06$ keV, which is now consistent with the expected value of 6.4 keV at a 90\% confidence limit. Moreover, as we shall see in the next subsections (in particular in \S \ref{sec:tor2}), we get good fits for the line feature adopting both toroidal models, in which the energy line is fixed to 6.4 keV, and a Compton shoulder is self-consistently calculated. Finally, since none of the previous observations found a such significant line energy shift, data with a better spectral resolution would be needed to assess the relevance of the one found here.
\subsection{Toroidal models}
\label{sec:tor}
\begin{figure}
\includegraphics[width=0.45\textwidth]{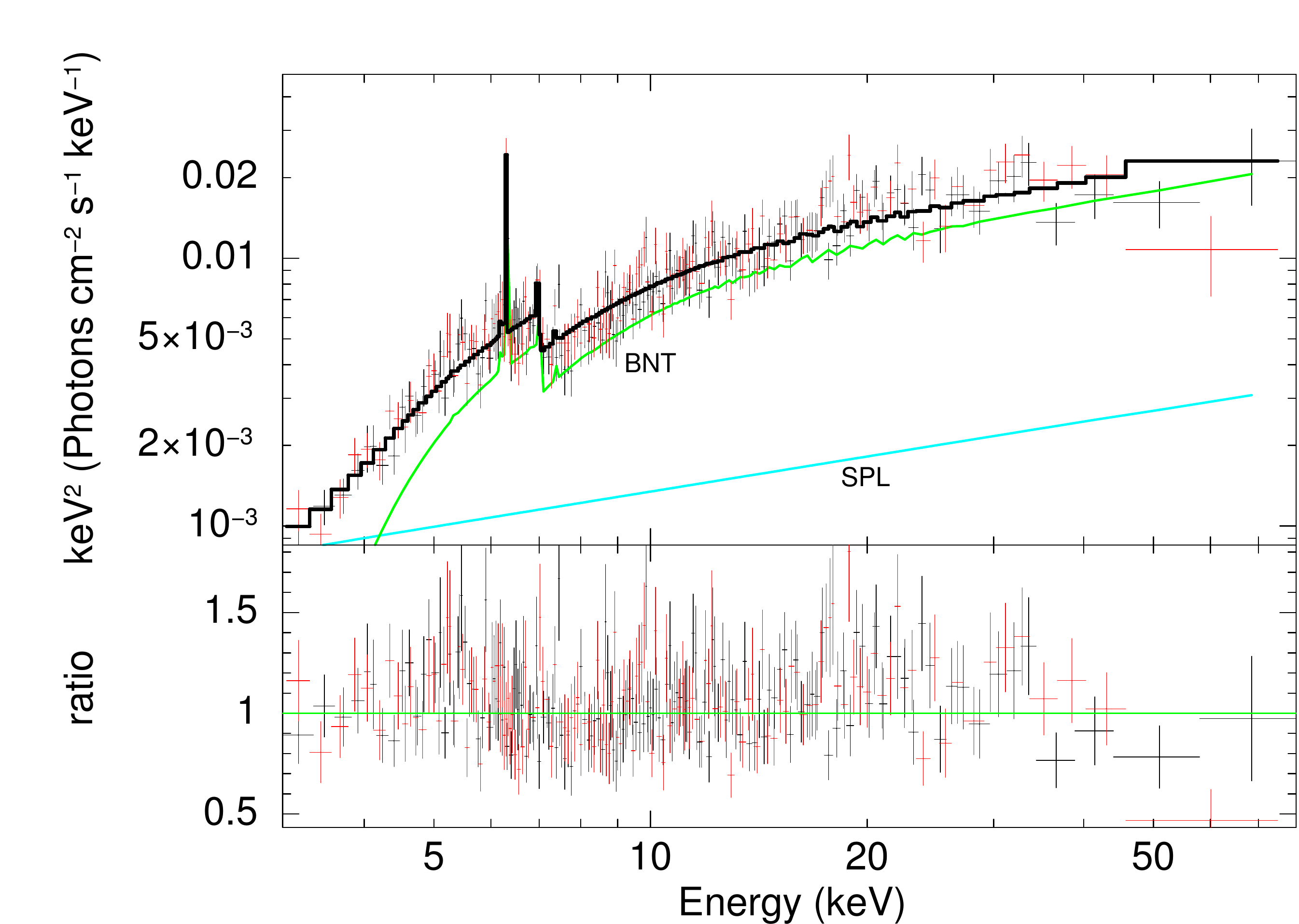}
\includegraphics[width=0.45\textwidth]{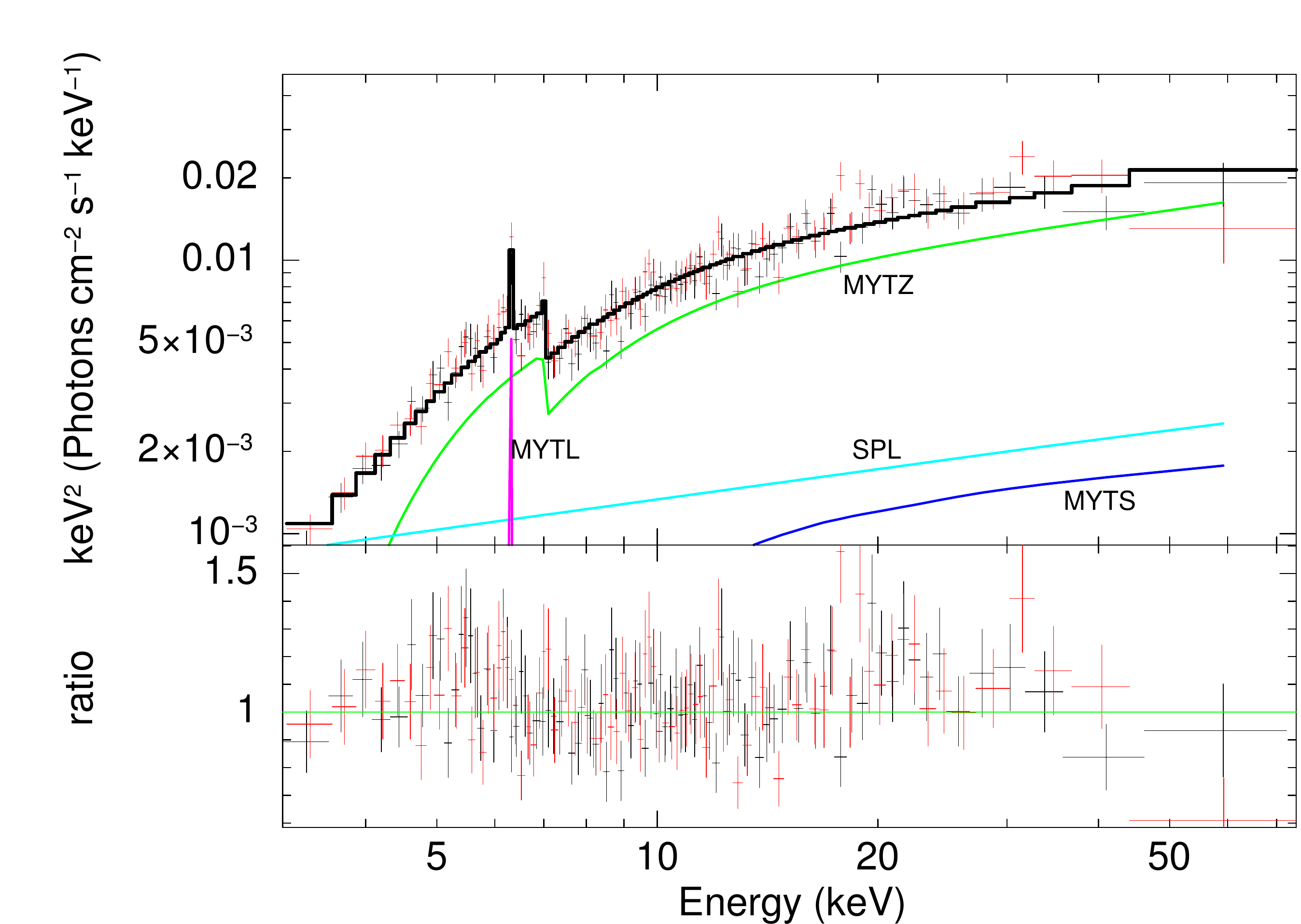}
\caption{Fit with with the BNTorus model (top) and the coupled MYTorus model (bottom) to the \nustar \ spectrum of Mrk\,1210 (FPMA in black, FPMB in red).}
\label{fig:toroidals}
\end{figure}

The high reflection value obtained applying the baseline model implies a more complicated geometry or some time delay between the components; at the same time, a reflection component alone is not able to successfully fit the spectrum ($\chi^{2}/\nu = 609/481$). Therefore, we explored more physically motivated models which can take into account both absorption and reflection in a self-consistent way. \newline
To this aim, we tried the BNTorus model \citepads{2011MNRAS.413.1206B} and the MYTorus model \citepads{2009MNRAS.397.1549M}. They have been recently developed based on Monte Carlo simulations, adopting a toroidal geometry for the material responsible for obscuration, scattering, and line fluorescence, all self-consistently treated. \newline
 We first applied the BNTorus model to the data, fixing the inclination angle of the torus (i.e., the parameter $\theta_{\rm inc}$; $\theta_{\rm inc} = 0$ describes a face-on view) to 87\degree. The fit was acceptable ($\chi^{2}/\nu = 517/481$), with $\Gamma = 1.57^{+ 0.10}_{- 0.06}$, $N_{\rm H} = 2.8^{+ 0.7}_{- 0.5} \times 10^{23}$ cm$^{-2}$, and half-opening angle of the torus $\theta_{ \rm tor} < 65\degree$, from which a covering factor of $>$ 0.4 is inferred. A very high fraction, $16^{+ 4}_{- 5}\%$, of the primary power law is scattered below 5 keV.  
 %Leaving the inclination angle free, the fit returns a slightly lower value ( $\theta_{\rm inc} \sim$ 63 \degree), but the parameter is unconstrained.
\newline Similarly, the MYTorus model in its ``coupled'', default mode (i.e. with a donut-shaped, smooth reprocessor with a covering factor of 0.5) gives an acceptable fit: $\chi^{2}/\nu = 512/480$, $\Gamma = 1.63 \pm 0.08$, $N_{\rm H} = (3.5 \pm 0.6) \times 10^{23}$ cm$^{-2}$, with a high fraction, $(12 \pm 3)\%$, of the primary power law scattered below 5 keV. However, using either torus model, the spectrum is still too peaky at $\sim 30-40$ keV (see Figure \ref{fig:toroidals}), and we conclude that they are not able to capture the spectral shape of Mrk\,1210 while the phenomenological, baseline model does. This suggests that the Mrk\,1210 torus contains Compton-thick material producing the pronounced Compton hump, but our line of sight does not pass the Compton-thick part in the \nustar\ observation.
\subsection{A physical picture}
\label{sec:tor2}
\begin{figure}
\centering
\includegraphics[width=0.45\textwidth]{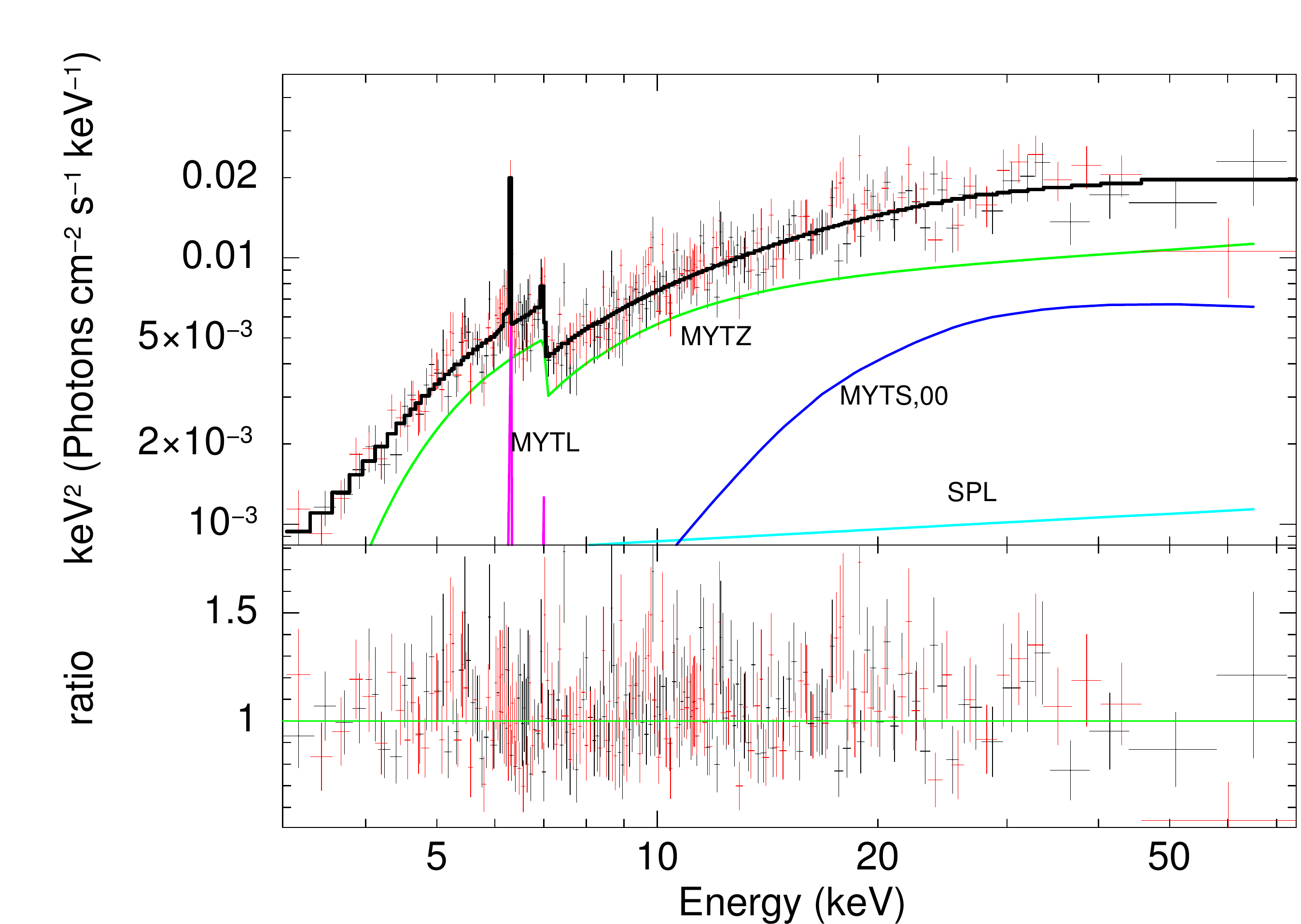}
\caption{Fit with the MYTorus decoupled model \eqref{eq:mytdec}, with normalizations and column densities decoupled (FPMA in black, FPMB in red). Fitting parameters are reported in Table \ref{table:specanalysis}.}
\label{fig:mytd_nhdecoupled}
\end{figure}

In the previous Section we found that the reflection component, modeled with a \texttt{pexrav} model, has a normalization a factor $\sim 2.5\times$ larger than the primary continuum from the AGN. Indeed, applying other self-consistent models based on Monte Carlo simulations with a classic toroidal geometry, we found worse fits with respect to that obtained with the baseline model, due to the presence of the reflection excess. 
\begin{comment}
We then added a \texttt{pexrav} model to our toroidal models. We fixed the reflection parameter to -1 and we linked the photon index to the torus model one. Both the BNTorus and the MYTorus models improved (from  $\chi^{2}/\nu = 517/481$ to  $\chi^{2}/\nu = 485/480$ and from  $\chi^{2}/\nu = 512/480$ to  $\chi^{2}/\nu = 486/479$, respectively; this implies that \texttt{pexrav} is significant at more than 99\% of confidence limit). See Figures \ref{fig:torus+pexrav} and \ref{fig:mytc+pexrav} for best fit models and data/model ratios. \newline
\end{comment}
\newline 
While the BNTorus model allows the user to fit for the torus opening angle and hence to get an estimate of the covering factor of the source, the three components of the MYTorus model can be decoupled (transmitted flux, reflected flux, and fluorescence lines) to simulate different -- and more complex -- geometries (we refer to \citeads{2012MNRAS.423.3360Y} for an extensive explanation of the decoupled version of the model, but see also \citeads{2014ApJ...787...61L} for a systematic application of the model on a sample of sources).
Briefly, together with the canonical components with a l.o.s. angle fixed at 90\degree (called front-scattered, with the subscript 90), the decoupled version includes an additional couple of scattered+line components seen ``face on'', i.e. with the l.o.s.  angle fixed to 0\degree (called back-scattered, with the subscript 00). The relative normalizations  between the 90\degree and 0\degree components (called $A_{\rm S90}$ and $A_{\rm S00}$, respectively) can be untied and left free in the fit. Additionally, one can also untie the column densities of the two scattered/reflected components. In XSPEC notation, this model is described as
\begin{equation}
\label{eq:mytdec}
\begin{split}
\mathrm{MYT\_DEC} =\overbrace{ \mathrm{constant}}^{\text{cross-normalization}}\times\overbrace{\mathrm{phabs}}^{\text{Gal. absorption}}\times \\
\times~ (\overbrace{ \mathrm{zpowerlw}\times \mathrm{MYTZ}}^{\text{absorption}} +\overbrace{\mathrm{MYT_{S,90}}~+ \mathrm{MYT_{L,90}}}^{\text{front-scattering}} +\\ + \overbrace{\mathrm{MYT_{S,00}} + \mathrm{MYT_{L,00}} }^{\text{back-scattering}}~ + \overbrace{ f_{\rm s}\times\mathrm{zpowerlw}}^{\text{soft component (SPL)}}).
\end{split}
\end{equation}
If we decouple the model (i.e., we leave free the constants $A_{\rm S90}$ and $A_{\rm S00}$), and allow the obscuring columns to be different (i.e., $N_{\rm H90} \neq N_{\rm H00}$), we get a good fit ($\chi^{2}/\nu = 489/479$, Figure \ref{fig:mytd_nhdecoupled}).  The front-scattered component vanishes (i.e.,  $A_{\rm S90} \rightarrow 0$), while the back-scattered component converges to the same normalization as the primary continuum (i.e.,  $A_{\rm S00} \rightarrow 1$, preserving the internal self-consistency of the model), and the clouds responsible for the reflection component are Compton-thick, with an optical depth $\tau \sim 2.7$ ($N_{\rm H} \sim 4 \times 10^{24}$ cm$^{-2}$). Moreover, the upper limit on the column density is unconstrained by the fit. In other words, the column density obscuring the l.o.s. is different from the column density of the clouds responsible of the back-scattered reflection. Fixing the normalizations of the back-scattered component and the primary power law to be equal as suggested by the fit (i.e., $A_{\rm S00} = 1$), does not change the result regarding the different column densities for the two components. On the other hand, a worse fit is obtained ($\chi^{2}/\nu = 505/479$, $\Delta\chi^2= 16$ for the same number of degrees of freedom) if the reflection component is front-scattered, i.e., if we simply decouple the column densities from the default configuration of MYTorus. Also in this case, the data require two different absorbing columns: a Compton-thin one obscuring the primary emission, and a thick one producing the reflection. The parameters of the models used for fitting (baseline, toroidal, MYTorus decoupled) are shown in Table \ref{table:specanalysis}.
\begin{comment}
\begin{figure}
\centering
\includegraphics[width=0.45\textwidth]{torus+pexrav}
\caption{Fit with the BNTorus + \texttt{pexrav} model to the \nustar \ spectrum of Mrk\,1210 (FPMA in black, FPMB in red).}
\label{fig:torus+pexrav}
\end{figure}
\begin{figure}
\centering
\includegraphics[width=0.45\textwidth]{mytc+pexrav}
\caption{Fit with the MYTorus (coupled) + \texttt{pexrav} model to the \nustar \ spectrum of Mrk\,1210 (FPMA in black, FPMB in red).}
\label{fig:mytc+pexrav}
\end{figure}
\end{comment}
\begin{table*}
\caption{Summary of spectral analysis.}             
\label{table:specanalysis}      
\centering          
\begin{tabular}{l c c c c c c c c c c} 
\hline\hline       
\noalign{\vskip 0.5mm}  
Parameter  & \multicolumn{3}{c}{Models} \\ \noalign{\vskip 0.5mm} 
 & BASELINE & MYT\tablefootmark{*} & MYT decoupled \\ 
\noalign{\vskip 0.5mm}  
\cline{2-4}                   
\noalign{\vskip 1mm}  
$\chi^2$/$\nu$ & 478/478 & 512/480 & 489/479\\ \noalign{\vskip 0.5mm} 
$\Gamma$ & 1.90$^{+ 0.11}_{- 0.12}$  & 1.63$^{+ 0.08}_{- 0.07}$ & 1.85 $\pm$ 0.12 \\ \noalign{\vskip 0.5mm} 
$N_{\rm H}$ [cm$^{-2}$] & 3.0$^{+ 0.7}_{- 0.6}$ $\times$ 10$^{23}$ & 3.5$^{+ 0.6}_{- 0.6}$ $\times$ 10$^{23}$ & 3.3$^{+ 0.8}_{- 0.7}$  $\times$ 10$^{23}$ \\ \noalign{\vskip 0.5mm} 
Norm transmitted comp @1 keV [ph cm$^{-2}$ s$^{-1}$ keV$^{-1}$] & 5.1$^{+ 1.3}_{- 1.0}$ $\times$ 10$^{-3}$ &  4.8$^{+ 1.3}_{- 1.1}$ $\times$ 10$^{-3}$ & 7.8$^{+ 3.8}_{- 2.6}$ $\times$ 10$^{-3}$  \\ \noalign{\vskip 1mm} 
Norm reflected comp @1 keV [ph cm$^{-2}$ s$^{-1}$ keV$^{-1}$]\tablefootmark{a} & 1.2$^{+ 0.9}_{- 0.5}$ $\times$ 10$^{-2}$  & 4.8 $\times$ 10$^{-3}$ (fixed) & 7.7$^{+ 5.1}_{- 3.1}$ $\times$ 10$^{-3}$ \\ \noalign{\vskip 0.5mm}
$N_{\rm H00}$ [cm$^{-2}$]\tablefootmark{b} &  &  & 4.0$^{+ u}_{- 1.6}$  $\times$ 10$^{24}$  \\ \noalign{\vskip 0.5mm}  
Line Energy [keV] & 6.28$^{+ 0.06}_{- 0.07}$  &  \\ \noalign{\vskip 0.5mm}
EW Line [eV] & 141 $\pm$ 50 & \\ \noalign{\vskip 0.5mm}
Norm line component (flux) [ph cm$^{-2}$ s$^{-1}$] & $(1.9 \pm 0.7) \times 10^{-5}$  &   \\ \noalign{\vskip 0.5mm}
$f_{\rm s}$ [\%] & 8$^{+ 5}_{- 6}$ & 12$^{+ 3}_{- 2}$ & 8$^{+ 3}_{- 2}$ \\ \noalign{\vskip 0.5mm}
$F_{2-10}$ [erg cm$^{-2}$ s$^{-1}$] & 7.6 $\times$ 10$^{-12}$ & 7.8 $\times$ 10$^{-12}$ & 7.7 $\times$ 10$^{-12}$  \\ \noalign{\vskip 0.5mm} 
$F_{10-40}$ [erg cm$^{-2}$ s$^{-1}$] & 3.0 $\times$ 10$^{-11}$ & 2.9 $\times$ 10$^{-11}$ & 3.0 $\times$ 10$^{-11}$  \\ \noalign{\vskip 0.5mm} 
$L^{\rm int}_{2-10}$ [erg s$^{-1}$] & 6.0 $\times$ 10$^{42}$ &  8.6 $\times$ 10$^{42}$ &  1.0 $\times$ 10$^{43}$   \\ \noalign{\vskip 0.5mm} 
$L^{\rm int}_{10-40}$ [erg s$^{-1}$] & 6.1 $\times$ 10$^{42}$ & 1.3 $\times$ 10$^{43}$ & 1.1 $\times$ 10$^{43}$  \\ \noalign{\vskip 0.5mm} 
FPMB/FPMA & 1.05 $\pm$ 0.03 & 1.05 $\pm$ 0.03 & 1.05 $\pm$ 0.03 \\ 
\noalign{\vskip 1mm}    
\hline              
\end{tabular}
\tablefoot{The values of the fluxes reported in the table are the observed ones, while those of the luminosities are intrinsic, i.e. deabsorbed.
\tablefoottext{*} {The results using a BNTorus model are the same within the uncertainties, and the main parameters of the fit are described in the text. We chose to show the MYTorus results to facilitate comparison between the coupled and decoupled cases. }
\tablefoottext{a} {The reflection component is the \texttt{pexrav} model in the first column. In the coupled configuration of MYTorus, absorption, fluorescence and reflection are self-consistently treated. For this reason, only one normalization is needed to describe all the components (second column). In the last column, the reflection is instead made up by the back-scattered MYTorus component, namely MYTS,00.}
\tablefoottext{b} {The column density is associated with the back-scattered reflection component in the decoupled MYTorus model, MYTS,00. }. }
\end{table*}
\subsection{Intrinsic luminosity}
Our models find an intrinsic 2--10 keV luminosity in the range $0.6-1 \times 10^{43}$ erg s$^{-1}$. It is interesting to compare these results with other commonly used proxies for the X-ray luminosity. Using the mid-infrared luminosity, the \textit{WISE} all-sky catalog \citepads{2010AJ....140.1868W, 2013wise.rept....1C} reports a $W3$ (12 $\mu m$) magnitude of $4.634\pm0.015$, which translates to a 12 micron luminosity $L_{12 \mu\text{m}} = 4.6 \times 10^{43}$ erg s$^{-1}$, with a $\sim$ 1.5\% statistical error, using the standard \textit{WISE} zeropoints. The $W1-W2$ color is $1.392\pm0.030$, suggesting that the mid-infrared luminosity is AGN-dominated, being the source above the color threshold of $W1-W2>0.8$ identified by \citetads{2012ApJ...753...30S}.
Then, from the mid-IR/X-ray relation \citepads{2009A&A...502..457G, 2015MNRAS.454..766A}, one would predict $\log(L_{2-10}[\text{erg/s}]) \sim 43.3$, which is consistent with our MYTorus luminosity. \newline
We can also use the optical [OIII] emission line to provide another independent estimate of the X-ray luminosity. Koss et al. (submitted) report a dust reddening-corrected [OIII]$\lambda$5007 flux of $F_{\rm[OIII]}^{\rm c} \sim 8.03 \times 10^{-13}$ erg cm$^{-2}$ s$^{-1}$, from which we derive a luminosity of $L_{\rm[OIII]} \sim 3.0 \times 10^{41}$ erg s$^{-1}$. Using the relationship for Seyfert galaxies between $L_{2-10}$ and $L_{\rm[OIII]}$ with a scatter of 0.5 dex presented in \citetads{2015MNRAS.454.3622B}, we predict $\log(L_{2-10}[\text{erg/s}]) \sim 43.3$, consistent with the luminosity derived from the infrared and with that obtained by our spectral analysis. These values suggest that the luminosity derived by the ``baseline'' model is likely underestimated of a factor $\sim$ 2, due to the phenomenological combination of the absorbed power law and infinite slab reflection models. The physically motivated decoupled MYTorus model alleviates this problem and allows a more reliable estimate of the intrinsic luminosity of the source. 
\section{Long-term behavior and discussion}
\label{sec:longterm}
\begin{figure*}
\centering
\includegraphics[width=\textwidth]{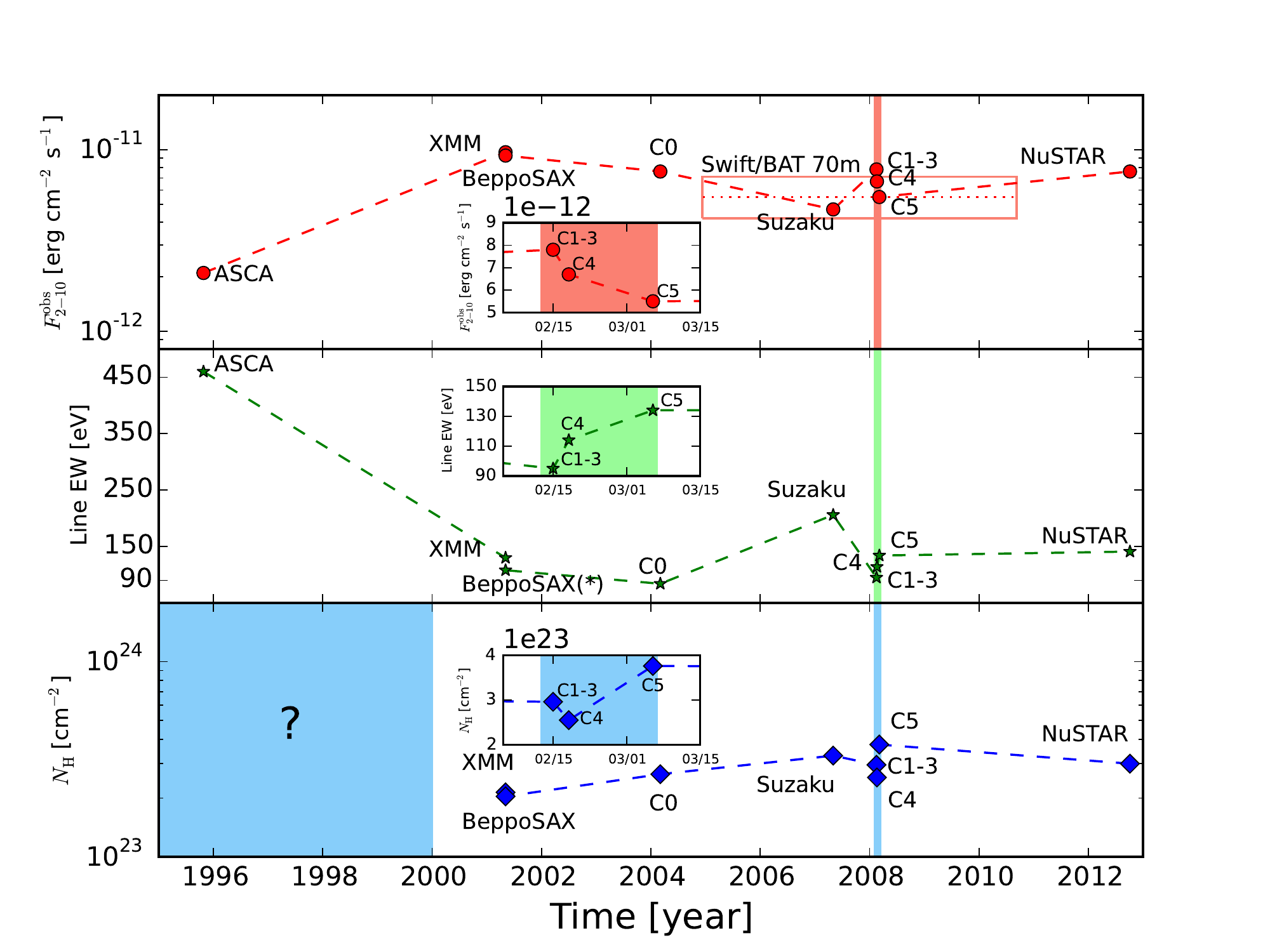}
\caption{Observed flux in the 2$-$10 keV band of Mrk\,1210 (top panel), Fe K$\alpha$ line equivalent width (middle panel) and column density (bottom panel) as a function of time. The inset panels show the zoomed-in region of the five \chandra\ observations during 2008. (*) We report here the equivalent width of model 1 instead of that of model 2 in \citetads{2004PASJ...56..425O}. The two values are consistent within the uncertainties, and we adopt the more constrained one for clarity.}
\label{fig:lc}
\end{figure*}
In order to understand the time dependent variation of the X-ray spectrum, we compiled the observed 2$-$10 keV fluxes, iron line equivalent widths, and column densities from the literature together with results of this work, to have a global picture of the behavior of the source. We choose to use our ``baseline'' model parameters (shown in Table \ref{table:specanalysis}) in order to compare directly with the results from previous papers based on the same phenomenological models. To do so, we reduced the data of the six \chandra\ observations with standard procedures. See the Appendix for the analysis of the 2004 observation (hereafter ``C0''), while for the other five observations during 2008 (namely, ``C1-5'') the best-fitting models of \citetads{2010MNRAS.406L..20R}, analogous to our baseline model, were applied. We also note that Mrk\,1210 is bright enough to be detected by the Burst Alert Telescope (BAT) onboard \textit{Swift}, and it is indeed present in the \textit{Swift}/BAT 70 month catalog \citepads{2013ApJS..207...19B}. As can be easily seen from the top panel of Figure \ref{fig:lc}, the source was in a low-flux state during the \textit{ASCA} observation. It has been seen in a high-flux state since then, with different observatories. 
The equivalent widths of the iron line (Figure \ref{fig:lc}, middle panel) reflect this trend. From the bottom panel of Figure \ref{fig:lc} it is clear that the column density of the source has been around $2-4 \times 10^{23}$ cm$^{-2}$ since 2001, while the \textit{ASCA} data do not help in shedding light on the column density in 1995. Both \citetads{2000ApJ...542..175A} and \citetads{2002A&A...388..787G} found a Compton-thin obscuration applying a pure transmission model to the \textit{ASCA} data, despite getting a worse fit with respect to a reflection-dominated scenario. On the other hand, the absorption-corrected 2$-$10 keV luminosity reported by \citetads{2000ApJ...542..175A} is a factor of $\sim$ 5 lower than the later intrinsic luminosities, but the quality of the data does not allow a robust estimate of the intrinsic luminosity, nor a robust detection of the transmitted continuum, leaving total degeneracy between heavily absorbed and intrinsically faint primary emission scenarios. In the following, both the flux change (\S \ref{sec:fluxchange}) and the eclipsing scenario (\S \ref{sec:eclipse}) are discussed.

\begin{table}
\caption{\textit{ASCA} best fit parameters compared with \nustar\ \texttt{pexrav}+\texttt{zgauss} only model.}
\label{tab:ascanustar}
\centering
\begin{tabular}{lcc}
\hline\hline
\noalign{\vskip 0.5mm}   
Parameter & \textit{ASCA} & \nustar\ (pexrav+line)  \\ \noalign{\vskip 0.5mm} 
\noalign{\vskip 0.5mm}  
\hline
\noalign{\vskip 1mm} 
$\Gamma$ & $1.95^{+ 0.45}_{-0.40}$ & $1.90^{+ 0.11}_{- 0.12}$ \\ \noalign{\vskip 0.5mm}  
Line EW [eV] & $460 \pm 210$ & 484  \\ \noalign{\vskip 0.5mm}  
$F_{2-10}$ [cgs] & $1.8 \times 10^{-12}$ & $2.3 \times 10^{-12}$  \\ \noalign{\vskip 0.5mm}  
$L_{2-10}^{\rm int}$ [cgs] & $2.3 \times 10^{42}$ & $0.9 \times 10^{42}$  \\
\noalign{\vskip 1mm}  
\hline
\end{tabular}
\tablefoot{The  \nustar\ line equivalent width is calculated with respect to the reflection continuum.}
\end{table}

\subsection{Change in intrinsic luminosity}
\label{sec:fluxchange}
If we suppose that Mrk\,1210 was Compton-thin also during the \textit{ASCA} observation, but with the intrinsic emission shut off, the reflection-dominated spectrum was entirely due to an echo of a previous high-flux state. \newline Thanks to the high-quality \nustar\ spectrum, we are able to disentangle the intrinsic power law and the reflection component. If we shut off the intrinsic component in our \nustar \ data, leaving only the \texttt{pexrav} model and the line component (i.e., a reflection dominated model), we find an observed flux and line equivalent width consistent with the values reported by \citetads{2000ApJ...542..175A}, albeit uncertainties on these values cannot be reliably computed, being a reflection dominated scenario highly disfavored by the \nustar\ data, as already discussed in \S \ref{sec:spec} (see Table \ref{tab:ascanustar}). \newline
Moreover, from the \nustar\ best fit baseline model, we can easily adjust the fit parameters, and vary only the normalization of the intrinsic continuum, to recover all the subsequent states of Mrk\,1210, which are similar to the \nustar \ one. This means that keeping a reflection component constant and varying only the intrinsic continuum, together with minor -- a factor $\sim$ 2, but required -- column density variations on shorter timescales (as noted by \citeads{2010MNRAS.406L..20R}), all the observations of the Phoenix galaxy during the last $\sim$ 20 years can be explained.
\begin{figure}
\centering
\includegraphics[width=0.5\textwidth]{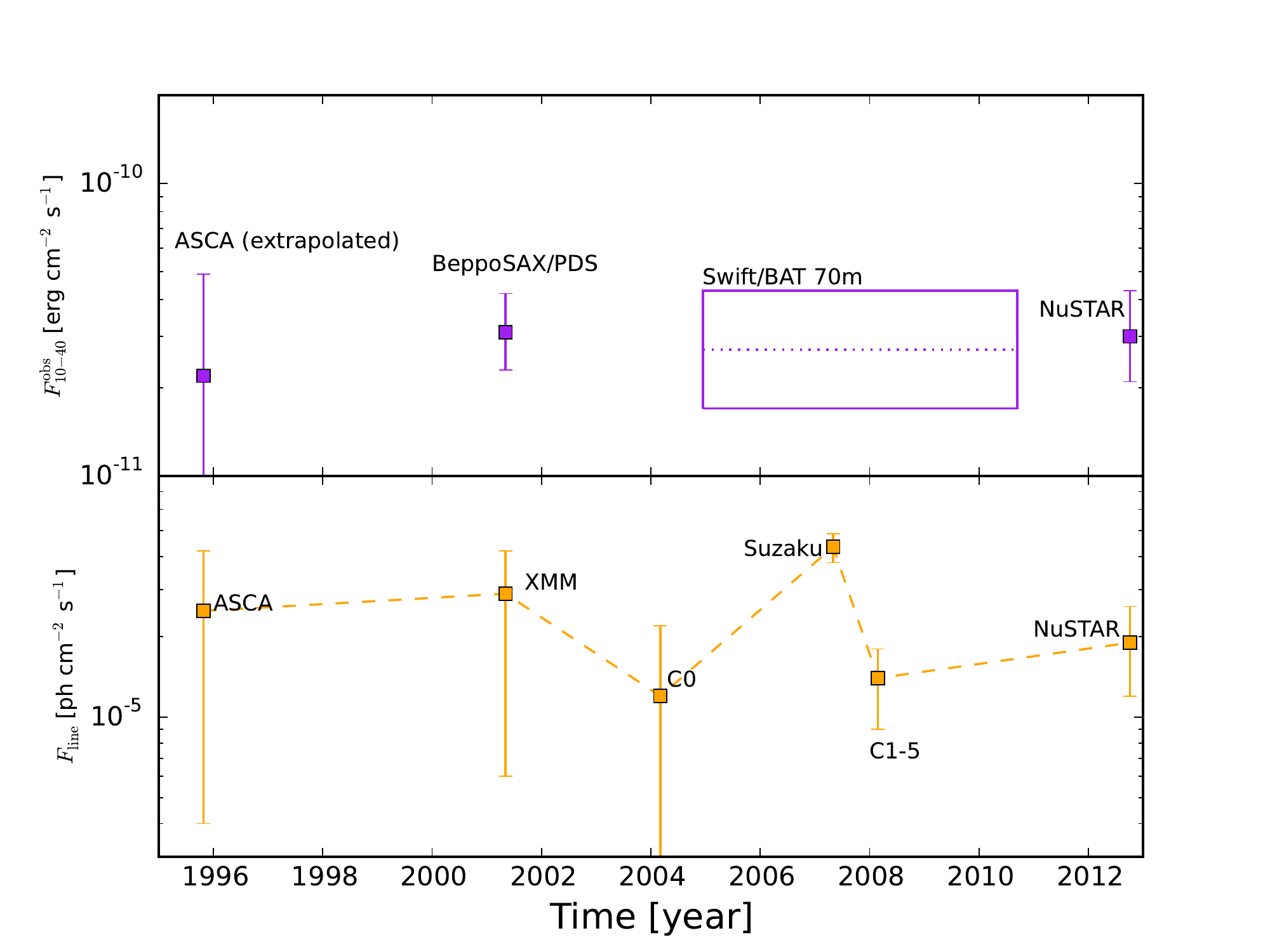}
\caption{Observed 10$-$40 keV flux (top panel) and iron line flux (bottom panel) of Mrk\,1210, as a function of time.}
\label{fig:refl_trend}
\end{figure}
\newline
As a further check, the 10--40 keV observed flux and line flux are plotted in Figure \ref{fig:refl_trend}. In the top panel, we computed the 10--40 keV flux, which is the band where the reflection component is thought to dominate, from \textit{BeppoSAX}/PDS\footnote{adapted from http://www.asdc.asi.it/bepposax/nfiarchive/reproc/\\5125800200/html/index.html}, \textit{Swift}/BAT 70 month catalog\footnote{the 14$-$195 keV flux was converted to a 10$-$40 keV using the \texttt{dummyrsp} command in XSPEC.}, and \nustar. We also extrapolated the \textit{ASCA} 10$-$40 keV flux, using the same procedure of \textit{Swift}/BAT. The bottom panel shows the line flux for the iron line component, measured by different instruments. Both panels seem to confirm the constancy of the reflection through years, with a hint of a slightly enhanced line during the \suzaku\ observation. 
\newline If this is the correct picture, Mrk\,1210 turned on between 1995 and 2001, and stayed more or less constant since then. The reflection component, though, still has to adjust and respond to this change, since it can be considered constant between all the observations. A simple light-travel argument can then put constraints on the distance between the central source and the reflector. For an edge-on view of the system, $D = c\Delta t/2$, and $\Delta t > 12 - 17$ yr, depending on when the source switched on (between the beginning of 1996 and the beginning of 2001). In this case, the distance between the source and the reflector is $D > (1.9 - 2.6)$ pc. First a fading, and then an increase in reflection is then expected to occur in the future, and monitoring of the source is the only way to keep track of its changes. Monitoring could also shed new light on the role of the X-ray absorber, which varies on much shorter temporal scales and seems to be associated with broad line region (BLR) clouds \citepads[see]{2010MNRAS.406L..20R}. If this is the case, the physical interpretation of our result applying the decoupled MYTorus model is straightforward: the l.o.s. is intercepting the variable (and presumably compact) Compton-thin absorber, while the reflector is located on larger scales and the photons reflecting back to our l.o.s. are ``seeing'' a Compton-thick column of gas.  \citetads{2010MNRAS.406L..20R} suggest that the absorber is a BLR cloud, assuming a central black hole (BH) mass of $5-7 \times 10^7~ M_\odot$. In this picture, a second ``screen'' is required to obscure the broad lines in the optical spectrum, which is instead likely showing an outflow, resulting in the classification of Mrk\,1210 as a Seyfert\,2 \citepads{2007A&A...463..445M}. 
\begin{comment}We suggest an alternative interpretation. We note that, properly rescaling the final result of \citetads{2010MNRAS.406L..20R}, we obtain that the obscuring clouds are located $\sim M_7(210 \text{km/s}/v_{\rm K})^2$ pc away from the central engine (where $M_7$ is the black hole mass in units of $10^7 M_\odot$ and $v_{\rm K}$ is the Keplerian velocity), suggesting that the reflector and the absorber could be the same, axisymmetric, structure (i.e. the torus) if the mass of the black hole is $10^7 M_\odot$. In this picture, the l.o.s. is intercepting the edge of the near side of the torus (as also suggested by the fit with the BNTorus model in \S \ref{sec:spec}), which obscures the nucleus (and the BLR) causing the observed column density variability, while \end{comment} 
However, if the reflection component comes from photons reflecting off the far inner wall of the torus, delayed with respect to the intrinsic emission, a rather distant torus edge is needed with respect to the dust sublimation radius $R_{\rm sub}$, which is usually thought to mark the inner wall of the torus, and in this case is $\sim 0.2$ pc (\citeads{2009A&A...502..457G} and references therein, and adopting a bolometric correction of 20, \citeads{2012MNRAS.425..623L}).

\subsection{Eclipsing event}
\label{sec:eclipse}
The \textit{ASCA} data cannot distinguish between a Compton-thick eclipsing event and an intrinsic low-flux state of the central engine. A random pass of a Compton-thick cloud (or an alignment of clouds along the l.o.s.) during the \textit{ASCA} observation is consistent with the known clumpiness of the gas surrounding obscured AGN. During the two observations by \textit{ASCA} in 1995, separated by 25 days, Mrk\,1210 was in a low-flux state. In an eclipsing scenario, this means that the putative eclipsing event lasted for at least 25 days. Assuming Keplerian motion of the clump, the linear size eclipsed by a moving Compton-thick cloud in a given time interval can be written as $s = 4.5 \times 10^{13}(M_7/D_{\rm pc})^{1/2}\Delta t_{25}$ cm, where $D_{\rm pc}$ is the distance from the center in parsecs, and $\Delta t_{25}$ is the eclipse time interval in units of 25 days. This implies that the cloud is rotating with a Keplerian velocity $v_{\rm K} \sim 210(M_7/D_{\rm pc})^{1/2}$ km/s. Finally, assuming a Compton-thick column density the average density of the clump is $\rho \sim 2 \times 10^{10} N_{\rm H,24}(D_{\rm pc}/M_7)^{1/2}/\Delta t_{25}$ cm$^{-3}$, where $N_{\rm H,24}$ is the column density in units of $10^{24}$ cm$^{-2}$. Interestingly, high-density clumps on a parsec-scale (or subparsec-scale) are observed in some systems showing water maser emission at 22 GHz. Mrk\,1210 is indeed one of them \citepads{1994ApJ...437L..99B}, even if the association of the maser spots to a particular geometry is still unclear and sensitive VLBI observations are needed to further investigate the nuclear environment. The maser activity could indeed be associated with a dusty Keplerian disk orbiting the SMBH, from which the most precise BH mass available to date could be derived \citepads[e.g,]{2011ApJ...727...20K}. This would allow a direct comparison of this object with other samples of disk masers \citepads{2015ApJ...810...65P,2016A&A...589A..59M} and get clues on the X-ray absorber. On the other hand, the maser spots could either be associated with the outflow seen in the optical spectrum, invalidating any possible BH mass estimate. Moreover, as reported by \citetads{1998ApJ...501...94S} and \citetads{2007A&A...463..445M}, Mrk\,1210 probably shows a recent circumnuclear starburst. If the masers are associated with the outflow responsible for the broad components of the optical lines, the Phoenix galaxy could be a very interesting and local laboratory to study the interplay between AGN activity and star formation.
\section{Conclusions}
\label{sec:conclusions}
We presented the first \nustar \ observation of Mrk\,1210, also known as the Phoenix galaxy. It is a long-studied object, considered to be part of the ``changing-look'' class of AGN. 
\nustar\ observed Mrk\,1210 obscured by a Compton-thin column of gas, like other instruments over the past 16 years. The data are also showing an enhanced reflection component, which requires two different columns of gas to be taken into account.
Many previous studies have suggested the presence of a variable X-ray absorber, inducing changes in absorbing column of a factor $\sim$ 2 on the timescales of days and weeks. We note that these short-term column changes, properly complemented by long-term intrinsic variability of the central engine, are able to explain the line equivalent widths and fluxes observed by different instruments during the last $\sim$ 20 years. If the low flux-state observed by \textit{ASCA} in 1995 is due to an intrinsic fading of the engine, we note evidence that the reflection component has remained constant with time. In this scenario, we infer that the physical distance between the source and the reflector is of the order of at least 2 pc. First a drop, and then an enhancement of the reflection component are then expected to occur in the future, in response of the source ``low-flux state'' (seen by \textit{ASCA}) and soft X-ray ``awakening'' (i.e. the ``Phoenix effect'', \citeads{2002A&A...388..787G}) seen between 1996 and 2001.
On the other hand, if the low-flux state of Mrk\,1210 during the \textit{ASCA} observation can be ascribed to the presence of a Compton-thick cloud obscuring the l.o.s., the cloud can be identified with a maser-emitting clump on the sub-pc scale. Indeed, if the torus-like structure is clumpy, a random passage of an over-dense cloud along the l.o.s. is expected \citepads{2012ApJ...758...66W}. The frequency in which such events are to be expected could be evaluated with precise hydrodynamical simulations of the nuclear environment, but see \citetads{2014MNRAS.439.1403M} for a statistical study on this topic. \newline
Monitoring in the X-ray band is needed, either to detect the expected behavior of the reflection component and clarify the relation between the X-ray absorber and reflector, which seem to be on different spatial scales, or to possibly detect another low-flux state and further study the properties of the clumpy (sub)pc-scale absorber. To this extent, radio VLBI monitoring at 22 GHz would also be extremely useful to clarify the nature of the nuclear water maser activity.
\begin{comment}
\begin{table*}
\caption{Summary of spectral analysis.}             
\label{table:specanalysis}      
\centering          
\begin{tabular}{c c c c c} 
\hline\hline       
\noalign{\vskip 0.5mm}  
 $\chi^2/\nu$ & $\Gamma$ & $N_{\rm H}$ [cm$^{-2}$] & $F_{2-10}$ [erg cm$^{-2}$ s$^{-1}$]  & $L^{\rm int}_{2-10}$ [erg s$^{-1}$] \\           
\noalign{\vskip 1mm}  
 478/478 & 1.90$^{+ 0.11}_{- 0.12}$ & 3.0$^{+ 0.7}_{- 0.6}$ $\times$ 10$^{23}$ & 7.6 $\times$ 10$^{-12}$ &6.0 $\times$ 10$^{42}$ \\ \noalign{\vskip 1mm}    
\hline              
\end{tabular}
\end{table*}
\end{comment}
\begin{acknowledgements}
We thank the anonymous referee for useful suggestions that helped to improve the paper. \newline
This work was supported under NASA Contract NNG08FD60C, and it made use of data from the \nustar \ mission, a  project  led  by  the  California  Institute  of  Technology, managed by the Jet Propulsion Laboratory, and funded by the National Aeronautics and Space Administration. We thank the \nustar \ Operations, Software, and Calibration teams for support  with  the  execution  and  analysis  of  these  observations. This  research  made  use  of  the \nustar \ Data  Analysis Software  (NuSTARDAS)  jointly  developed  by  the  ASI  Science  Data  Center  (ASDC,  Italy)  and  the  California  Institute of Technology (USA). This research has also made use of data obtained from the \chandra\ Data Archive and the \chandra\ Source Catalog, and software provided by the \chandra\ X-ray Center (CXC). \newline 
A.\,M., A.\,C., and S.\,P. acknowledge support  from the ASI/INAF  grant  I/037/12/0-011/13. \newline
M.\,B. acknowledges support from NASA Headquarters under the NASA Earth and Space Science Fellowship Program, grant NNX14AQ07H. \newline
P.\,G. and P.\,B. thank STFC for support (grant ST/J003697/2). \newline
S.\,L.\,M. is supported by an appointment to the NASA Postdoctoral Program at the NASA Goddard Space Flight Center, administered by Universities Space Research Association under contract with NASA. \newline
We acknowledge support from NASA NuSTAR A01 Award NNX15AV27G (F.\,E.\,B.), CONICYT-Chile grants Basal-CATA PFB-06/2007 (F.\,E.\,B., C.\,R.), FONDECYT Regular 1141218 (F.\,E.\,B.,C.\,R.), "EMBIGGEN" Anillo ACT1101 (F.\,E.\,B.,C.\,R.), the China-CONICYT fund (C.\,R.), and the Ministry of Economy, Development, and Tourism's Millennium Science Initiative through grant IC120009, awarded to The Millennium Institute of Astrophysics, MAS (F.\,E.\,B.).
\end{acknowledgements}
%-------------------------------------------------------------------
\bibliographystyle{aa} % style aa.bst 
\bibliography{mrk1210} % your references Yourfile.bib

\begin{thebibliography}{39}
\expandafter\ifx\csname natexlab\endcsname\relax\def\natexlab#1{#1}\fi

\bibitem[{{Arnaud}(1996)}]{1996ASPC..101...17A}
{Arnaud}, K.~A. 1996, in Astronomical Society of the Pacific Conference Series,
  Vol. 101, Astronomical Data Analysis Software and Systems V, ed. G.~H.
  {Jacoby} \& J.~{Barnes}, 17

\bibitem[{{Asmus} {et~al.}(2015){Asmus}, {Gandhi}, {H{\"o}nig}, {Smette}, \&
  {Duschl}}]{2015MNRAS.454..766A}
{Asmus}, D., {Gandhi}, P., {H{\"o}nig}, S.~F., {Smette}, A., \& {Duschl}, W.~J.
  2015, \mnras, 454, 766

\bibitem[{{Awaki} {et~al.}(2000){Awaki}, {Ueno}, {Taniguchi}, \&
  {Weaver}}]{2000ApJ...542..175A}
{Awaki}, H., {Ueno}, S., {Taniguchi}, Y., \& {Weaver}, K.~A. 2000, \apj, 542,
  175

\bibitem[{{Baumgartner} {et~al.}(2013){Baumgartner}, {Tueller}, {Markwardt},
  {Skinner}, {Barthelmy}, {Mushotzky}, {Evans}, \&
  {Gehrels}}]{2013ApJS..207...19B}
{Baumgartner}, W.~H., {Tueller}, J., {Markwardt}, C.~B., {et~al.} 2013, \apjs,
  207, 19

\bibitem[{{Berney} {et~al.}(2015){Berney}, {Koss}, {Trakhtenbrot}, {Ricci},
  {Lamperti}, {Schawinski}, {Balokovi{\'c}}, {Crenshaw}, {Fischer}, {Gehrels},
  {Harrison}, {Hashimoto}, {Ichikawa}, {Mushotzky}, {Oh}, {Stern}, {Treister},
  {Ueda}, {Veilleux}, \& {Winter}}]{2015MNRAS.454.3622B}
{Berney}, S., {Koss}, M., {Trakhtenbrot}, B., {et~al.} 2015, \mnras, 454, 3622

\bibitem[{{Bianchi} \& {Guainazzi}(2007)}]{2007AIPC..924..822B}
{Bianchi}, S. \& {Guainazzi}, M. 2007, in American Institute of Physics
  Conference Series, Vol. 924, The Multicolored Landscape of Compact Objects
  and Their Explosive Origins, ed. T.~{di Salvo}, G.~L. {Israel},
  L.~{Piersant}, L.~{Burderi}, G.~{Matt}, A.~{Tornambe}, \& M.~T. {Menna},
  822--829

\bibitem[{{Bianchi} {et~al.}(2005){Bianchi}, {Guainazzi}, {Matt}, {Chiaberge},
  {Iwasawa}, {Fiore}, \& {Maiolino}}]{2005A&A...442..185B}
{Bianchi}, S., {Guainazzi}, M., {Matt}, G., {et~al.} 2005, \aap, 442, 185

\bibitem[{{Braatz} {et~al.}(1994){Braatz}, {Wilson}, \&
  {Henkel}}]{1994ApJ...437L..99B}
{Braatz}, J.~A., {Wilson}, A.~S., \& {Henkel}, C. 1994, \apjl, 437, L99

\bibitem[{{Brightman} \& {Nandra}(2011)}]{2011MNRAS.413.1206B}
{Brightman}, M. \& {Nandra}, K. 2011, \mnras, 413, 1206

\bibitem[{{Cutri} {et~al.}(2013){Cutri}, {Wright}, {Conrow}, {Fowler},
  {Eisenhardt}, {Grillmair}, {Kirkpatrick}, {Masci}, {McCallon}, {Wheelock},
  {Fajardo-Acosta}, {Yan}, {Benford}, {Harbut}, {Jarrett}, {Lake}, {Leisawitz},
  {Ressler}, {Stanford}, {Tsai}, {Liu}, {Helou}, {Mainzer}, {Gettings},
  {Gonzalez}, {Hoffman}, {Marsh}, {Padgett}, {Skrutskie}, {Beck}, {Papin}, \&
  {Wittman}}]{2013wise.rept....1C}
{Cutri}, R.~M., {Wright}, E.~L., {Conrow}, T., {et~al.} 2013, {Explanatory
  Supplement to the AllWISE Data Release Products}, Tech. rep.

\bibitem[{{Gandhi} {et~al.}(2009){Gandhi}, {Horst}, {Smette}, {H{\"o}nig},
  {Comastri}, {Gilli}, {Vignali}, \& {Duschl}}]{2009A&A...502..457G}
{Gandhi}, P., {Horst}, H., {Smette}, A., {et~al.} 2009, \aap, 502, 457

\bibitem[{{Gilli} {et~al.}(2000){Gilli}, {Maiolino}, {Marconi}, {Risaliti},
  {Dadina}, {Weaver}, \& {Colbert}}]{2000A&A...355..485G}
{Gilli}, R., {Maiolino}, R., {Marconi}, A., {et~al.} 2000, \aap, 355, 485

\bibitem[{{Guainazzi}(2002)}]{2002MNRAS.329L..13G}
{Guainazzi}, M. 2002, \mnras, 329, L13

\bibitem[{{Guainazzi} {et~al.}(2002){Guainazzi}, {Matt}, {Fiore}, \&
  {Perola}}]{2002A&A...388..787G}
{Guainazzi}, M., {Matt}, G., {Fiore}, F., \& {Perola}, G.~C. 2002, \aap, 388,
  787

\bibitem[{{Harrison} {et~al.}(2013){Harrison}, {Craig}, {Christensen},
  {Hailey}, {Zhang}, {Boggs}, {Stern}, {Cook}, {Forster}, {Giommi},
  {Grefenstette}, {Kim}, {Kitaguchi}, {Koglin}, {Madsen}, {Mao}, {Miyasaka},
  {Mori}, {Perri}, {Pivovaroff}, {Puccetti}, {Rana}, {Westergaard}, {Willis},
  {Zoglauer}, {An}, {Bachetti}, {Barri{\`e}re}, {Bellm}, {Bhalerao},
  {Brejnholt}, {Fuerst}, {Liebe}, {Markwardt}, {Nynka}, {Vogel}, {Walton},
  {Wik}, {Alexander}, {Cominsky}, {Hornschemeier}, {Hornstrup}, {Kaspi},
  {Madejski}, {Matt}, {Molendi}, {Smith}, {Tomsick}, {Ajello}, {Ballantyne},
  {Balokovi{\'c}}, {Barret}, {Bauer}, {Blandford}, {Brandt}, {Brenneman},
  {Chiang}, {Chakrabarty}, {Chenevez}, {Comastri}, {Dufour}, {Elvis}, {Fabian},
  {Farrah}, {Fryer}, {Gotthelf}, {Grindlay}, {Helfand}, {Krivonos}, {Meier},
  {Miller}, {Natalucci}, {Ogle}, {Ofek}, {Ptak}, {Reynolds}, {Rigby},
  {Tagliaferri}, {Thorsett}, {Treister}, \& {Urry}}]{2013ApJ...770..103H}
{Harrison}, F.~A., {Craig}, W.~W., {Christensen}, F.~E., {et~al.} 2013, \apj,
  770, 103

\bibitem[{{Kalberla} {et~al.}(2005){Kalberla}, {Burton}, {Hartmann}, {Arnal},
  {Bajaja}, {Morras}, \& {P{\"o}ppel}}]{2005A&A...440..775K}
{Kalberla}, P.~M.~W., {Burton}, W.~B., {Hartmann}, D., {et~al.} 2005, \aap,
  440, 775

\bibitem[{{Kuo} {et~al.}(2011){Kuo}, {Braatz}, {Condon}, {Impellizzeri}, {Lo},
  {Zaw}, {Schenker}, {Henkel}, {Reid}, \& {Greene}}]{2011ApJ...727...20K}
{Kuo}, C.~Y., {Braatz}, J.~A., {Condon}, J.~J., {et~al.} 2011, \apj, 727, 20

\bibitem[{{LaMassa} {et~al.}(2014){LaMassa}, {Yaqoob}, {Ptak}, {Jia},
  {Heckman}, {Gandhi}, \& {Meg Urry}}]{2014ApJ...787...61L}
{LaMassa}, S.~M., {Yaqoob}, T., {Ptak}, A.~F., {et~al.} 2014, \apj, 787, 61

\bibitem[{{Lusso} {et~al.}(2012){Lusso}, {Comastri}, {Simmons}, {Mignoli},
  {Zamorani}, {Vignali}, {Brusa}, {Shankar}, {Lutz}, {Trump}, {Maiolino},
  {Gilli}, {Bolzonella}, {Puccetti}, {Salvato}, {Impey}, {Civano}, {Elvis},
  {Mainieri}, {Silverman}, {Koekemoer}, {Bongiorno}, {Merloni}, {Berta}, {Le
  Floc'h}, {Magnelli}, {Pozzi}, \& {Riguccini}}]{2012MNRAS.425..623L}
{Lusso}, E., {Comastri}, A., {Simmons}, B.~D., {et~al.} 2012, \mnras, 425, 623

\bibitem[{{Magdziarz} \& {Zdziarski}(1995)}]{1995MNRAS.273..837M}
{Magdziarz}, P. \& {Zdziarski}, A.~A. 1995, \mnras, 273, 837

\bibitem[{{Markowitz} {et~al.}(2014){Markowitz}, {Krumpe}, \&
  {Nikutta}}]{2014MNRAS.439.1403M}
{Markowitz}, A.~G., {Krumpe}, M., \& {Nikutta}, R. 2014, \mnras, 439, 1403

\bibitem[{{Masini} {et~al.}(2016){Masini}, {Comastri}, {Balokovi{\'c}}, {Zaw},
  {Puccetti}, {Ballantyne}, {Bauer}, {Boggs}, {Brandt}, {Brightman},
  {Christensen}, {Craig}, {Gandhi}, {Hailey}, {Harrison}, {Koss}, {Madejski},
  {Ricci}, {Rivers}, {Stern}, \& {Zhang}}]{2016A&A...589A..59M}
{Masini}, A., {Comastri}, A., {Balokovi{\'c}}, M., {et~al.} 2016, \aap, 589,
  A59

\bibitem[{{Matt} {et~al.}(2009){Matt}, {Bianchi}, {Awaki}, {Comastri},
  {Guainazzi}, {Iwasawa}, {Jimenez-Bailon}, \&
  {Nicastro}}]{2009A&A...496..653M}
{Matt}, G., {Bianchi}, S., {Awaki}, H., {et~al.} 2009, \aap, 496, 653

\bibitem[{{Mazzalay} \& {Rodr{\'{\i}}guez-Ardila}(2007)}]{2007A&A...463..445M}
{Mazzalay}, X. \& {Rodr{\'{\i}}guez-Ardila}, A. 2007, \aap, 463, 445

\bibitem[{{Murphy} \& {Yaqoob}(2009)}]{2009MNRAS.397.1549M}
{Murphy}, K.~D. \& {Yaqoob}, T. 2009, \mnras, 397, 1549

\bibitem[{{Ohno} {et~al.}(2004){Ohno}, {Fukazawa}, \&
  {Iyomoto}}]{2004PASJ...56..425O}
{Ohno}, M., {Fukazawa}, Y., \& {Iyomoto}, N. 2004, \pasj, 56, 425

\bibitem[{{Pesce} {et~al.}(2015){Pesce}, {Braatz}, {Condon}, {Gao}, {Henkel},
  {Litzinger}, {Lo}, \& {Reid}}]{2015ApJ...810...65P}
{Pesce}, D.~W., {Braatz}, J.~A., {Condon}, J.~J., {et~al.} 2015, \apj, 810, 65

\bibitem[{{Piconcelli} {et~al.}(2007){Piconcelli}, {Bianchi}, {Guainazzi},
  {Fiore}, \& {Chiaberge}}]{2007A&A...466..855P}
{Piconcelli}, E., {Bianchi}, S., {Guainazzi}, M., {Fiore}, F., \& {Chiaberge},
  M. 2007, \aap, 466, 855

\bibitem[{{Risaliti} {et~al.}(2010){Risaliti}, {Elvis}, {Bianchi}, \&
  {Matt}}]{2010MNRAS.406L..20R}
{Risaliti}, G., {Elvis}, M., {Bianchi}, S., \& {Matt}, G. 2010, \mnras, 406,
  L20

\bibitem[{{Risaliti} {et~al.}(2005){Risaliti}, {Elvis}, {Fabbiano}, {Baldi}, \&
  {Zezas}}]{2005ApJ...623L..93R}
{Risaliti}, G., {Elvis}, M., {Fabbiano}, G., {Baldi}, A., \& {Zezas}, A. 2005,
  \apjl, 623, L93

\bibitem[{{Rivers} {et~al.}(2015{\natexlab{a}}){Rivers}, {Balokovi{\'c}},
  {Ar{\'e}valo}, {Bauer}, {Boggs}, {Brandt}, {Brightman}, {Christensen},
  {Craig}, {Gandhi}, {Hailey}, {Harrison}, {Koss}, {Ricci}, {Stern}, {Walton},
  \& {Zhang}}]{2015ApJ...815...55R}
{Rivers}, E., {Balokovi{\'c}}, M., {Ar{\'e}valo}, P., {et~al.}
  2015{\natexlab{a}}, \apj, 815, 55

\bibitem[{{Rivers} {et~al.}(2015{\natexlab{b}}){Rivers}, {Risaliti}, {Walton},
  {Harrison}, {Ar{\'e}valo}, {Baur}, {Boggs}, {Brenneman}, {Brightman},
  {Christensen}, {Craig}, {F{\"u}rst}, {Hailey}, {Hickox}, {Marinucci},
  {Reeves}, {Stern}, \& {Zhang}}]{2015ApJ...804..107R}
{Rivers}, E., {Risaliti}, G., {Walton}, D.~J., {et~al.} 2015{\natexlab{b}},
  \apj, 804, 107

\bibitem[{{Stern} {et~al.}(2012){Stern}, {Assef}, {Benford}, {Blain}, {Cutri},
  {Dey}, {Eisenhardt}, {Griffith}, {Jarrett}, {Lake}, {Masci}, {Petty},
  {Stanford}, {Tsai}, {Wright}, {Yan}, {Harrison}, \&
  {Madsen}}]{2012ApJ...753...30S}
{Stern}, D., {Assef}, R.~J., {Benford}, D.~J., {et~al.} 2012, \apj, 753, 30

\bibitem[{{Storchi-Bergmann} {et~al.}(1998){Storchi-Bergmann}, {Fernandes}, \&
  {Schmitt}}]{1998ApJ...501...94S}
{Storchi-Bergmann}, T., {Fernandes}, R.~C., \& {Schmitt}, H.~R. 1998, \apj,
  501, 94

\bibitem[{{Wada}(2012)}]{2012ApJ...758...66W}
{Wada}, K. 2012, \apj, 758, 66

\bibitem[{{Walton} {et~al.}(2014){Walton}, {Risaliti}, {Harrison}, {Fabian},
  {Miller}, {Arevalo}, {Ballantyne}, {Boggs}, {Brenneman}, {Christensen},
  {Craig}, {Elvis}, {Fuerst}, {Gandhi}, {Grefenstette}, {Hailey}, {Kara},
  {Luo}, {Madsen}, {Marinucci}, {Matt}, {Parker}, {Reynolds}, {Rivers}, {Ross},
  {Stern}, \& {Zhang}}]{2014ApJ...788...76W}
{Walton}, D.~J., {Risaliti}, G., {Harrison}, F.~A., {et~al.} 2014, \apj, 788,
  76

\bibitem[{{Wright} {et~al.}(2010){Wright}, {Eisenhardt}, {Mainzer}, {Ressler},
  {Cutri}, {Jarrett}, {Kirkpatrick}, {Padgett}, {McMillan}, {Skrutskie},
  {Stanford}, {Cohen}, {Walker}, {Mather}, {Leisawitz}, {Gautier}, {McLean},
  {Benford}, {Lonsdale}, {Blain}, {Mendez}, {Irace}, {Duval}, {Liu}, {Royer},
  {Heinrichsen}, {Howard}, {Shannon}, {Kendall}, {Walsh}, {Larsen}, {Cardon},
  {Schick}, {Schwalm}, {Abid}, {Fabinsky}, {Naes}, \&
  {Tsai}}]{2010AJ....140.1868W}
{Wright}, E.~L., {Eisenhardt}, P.~R.~M., {Mainzer}, A.~K., {et~al.} 2010, \aj,
  140, 1868

\bibitem[{{Yaqoob}(1997)}]{1997ApJ...479..184Y}
{Yaqoob}, T. 1997, \apj, 479, 184

\bibitem[{{Yaqoob}(2012)}]{2012MNRAS.423.3360Y}
{Yaqoob}, T. 2012, \mnras, 423, 3360

\end{thebibliography}
\begin{appendix} %First online appendix
\section{Analysis of the \chandra\ 2004 (C0) observation}
\begin{figure}
\centering
\includegraphics[width=0.45\textwidth]{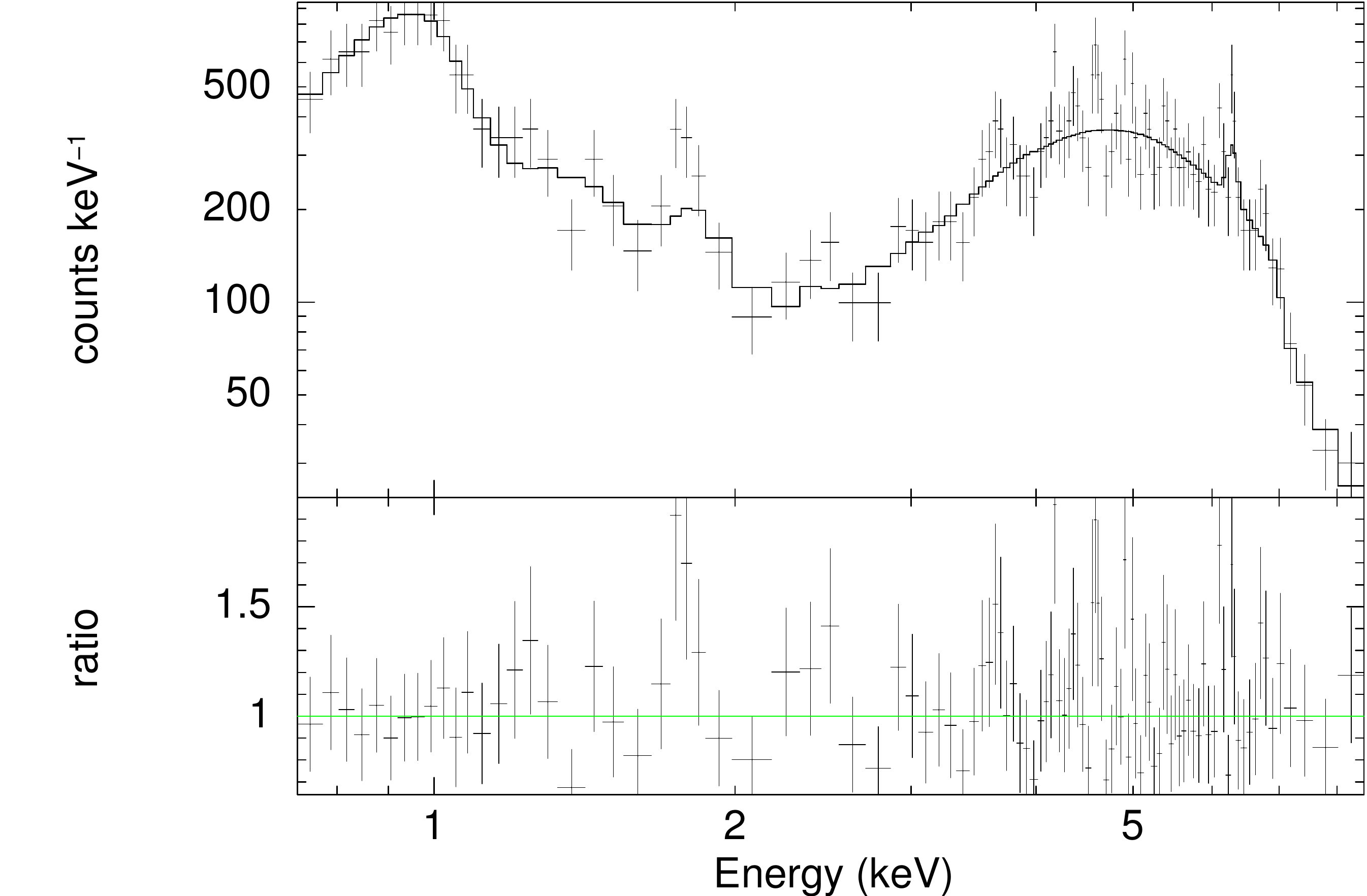}
\caption{Best fit and data/model ratio to the 2004 \chandra\ spectrum of Mrk\,1210.}
\label{fig:c0}
\end{figure}

We reduced the \chandra\ data and produced a grouped spectrum using the CIAO standard \texttt{chandra_repro} and \texttt{specextract} tasks. The source counts were extracted from a 3\arcsec-radius circular region, while the background was extracted from four 10\arcsec circular regions. Soft X-ray emission below 3 keV is clearly present in the total spectrum (Figure \ref{fig:c0}), and was fitted, following \citetads{2010MNRAS.406L..20R}, with an \texttt{apec+zpowerlw} model. The hard X-ray part of the spectrum was fitted with our ``baseline'' model, Equation \eqref{eq:baseline}. The fitting parameters are reported in Table \ref{tab:c0}. We note that the results are consistent with the \nustar\ observation presented in the paper.

\begin{table*}
\caption{C0 observation; best fitting parameters.}             
\label{tab:c0}      
\centering          
\begin{tabular}{l c} 
\hline\hline       
\noalign{\vskip 0.5mm}  
 Parameter & Value \\ \noalign{\vskip 1mm}
 $\chi^2/\nu$	& 83/95  \\ \noalign{\vskip 1mm}
\multicolumn{2}{c}{Hard component} \\ \noalign{\vskip 1mm}
$\Gamma$ & 1.56$^{+ 0.58}_{- 0.63}$ \\ \noalign{\vskip 0.5mm}
$N_{\rm H}$ [cm$^{-2}$] & 2.7$^{+ 0.8}_{- 1.0}$ $\times$ 10$^{23}$ \\ \noalign{\vskip 0.5mm}
Norm transmitted comp @1 keV [ph cm$^{-2}$ s$^{-1}$ keV$^{-1}$] & 2.2$^{+ 6.9}_{- 1.7}$ $\times$ 10$^{-3}$  \\ \noalign{\vskip 0.5mm}
Norm reflection comp @1 keV [ph cm$^{-2}$ s$^{-1}$ keV$^{-1}$] & 1.2$^{+1.0}_{- 1.0}$ $\times$ 10$^{-2}$ \\ \noalign{\vskip 0.5mm}
Line Energy [keV] & 6.36$^{+ 0.08}_{- 0.21}$  \\ \noalign{\vskip 0.5mm}
EW Line [eV] & 84 $^{+ 74}_{- 68}$  \\ \noalign{\vskip 0.5mm}
Norm line component (flux) [ph cm$^{-2}$ s$^{-1}$] & 1.2$^{+ 1.0}_{- 1.0}$ $\times$ 10$^{-5}$ \\ \noalign{\vskip 1mm}
\multicolumn{2}{c}{Soft component} \\ \noalign{\vskip 1mm}
$kT$ [keV] & 1.02$^{+ 0.16}_{- 0.09}$ \\ \noalign{\vskip 0.5mm}
Norm \texttt{apec} @1 keV [ph cm$^{-2}$ s$^{-1}$ keV$^{-1}$] & 4.3$^{+ 1.4}_{- 1.3}$ $\times$ 10$^{-5}$ \\ \noalign{\vskip 0.5mm}
$\Gamma_{\rm s}$ & 3.0$^{+ u}_{- l}$ \\ \noalign{\vskip 0.5mm}
Norm \texttt{zpowerlw} @1 keV [ph cm$^{-2}$ s$^{-1}$ keV$^{-1}$] & 2.9$^{+ 2.9}_{- 2.9}$ $\times$ 10$^{-5}$\\ \noalign{\vskip 0.5mm}
$F_{2-10}$ [erg cm$^{-2}$ s$^{-1}$]  & 7.6 $\times$ $10^{-12}$ \\
\noalign{\vskip 1mm}    
\hline              
\end{tabular}
\end{table*}
\end{appendix}
\end{document}